\documentclass[twocolumn,tighten]{aastex701} 
\hypersetup{linkcolor=red,citecolor=blue,filecolor=cyan,urlcolor=magenta}
\usepackage{amsmath}
\usepackage{natbib}
\usepackage{todonotes}
\usepackage{CJK}
\usepackage{xspace}
\usepackage[shortlabels]{enumitem}

\newcommand{\Gaia}{{\it Gaia}\xspace}

\newcommand{\kmsec}{\mbox{km~s$^{\rm -1}$}}

\newcommand{\msun}{\mbox{$M_{\odot}$}}

\newcommand{\RN}[1]{%
  \textup{\uppercase\expandafter{\romannumeral#1}}%
}

\newcommand{\bootes}{Bo\"otes}

\shorttitle{Binaries in UFD velocity dispersions}
\shortauthors{Ou et al.}
\begin{document}
\begin{CJK*}{UTF8}{gbsn}

\title{From observing strategies to velocity dispersion bias: forward modeling unresolved binaries in ultra-faint dwarf galaxies}

\author[0000-0002-4669-9967]{Xiaowei~Ou~(欧筱葳)} 
\altaffiliation{Galaxy Evolution and COsmology (GECO) Fellow}
\altaffiliation{CosmicAI Fellow}
\affiliation{%
Department of Astronomy, 
University of Virginia,
530 McCormick Rd, Charlottesville, VA 22904, USA}
\affiliation{The NSF-Simons AI Institute for Cosmic Origins, USA}
\email{Email:\ xwou@virginia.edu}

\author[0000-0002-6021-8760]{Andrew~B.~Pace}
\altaffiliation{Galaxy Evolution and COsmology (GECO) Fellow}
\affiliation{%
Department of Astronomy, 
University of Virginia,
530 McCormick Rd, Charlottesville, VA 22904, USA}
\email{Email:\ apace@virginia.edu}

\author[0000-0002-3204-1742]{Nitya~Kallivayalil}
\affiliation{%
Department of Astronomy, 
University of Virginia,
530 McCormick Rd, Charlottesville, VA 22904, USA}
\affiliation{The NSF-Simons AI Institute for Cosmic Origins, USA}
\email{Email:\ njk3r@virginia.edu}

\author[0000-0003-1379-6696]{Amery~Gration}
\affiliation{%
Department of Physics, 
University of Surrey,
GUILDFORD, GU2 7XH, United Kingdom}
\email{Email:\ a.gration@surrey.ac.uk}

\author[0000-0001-9061-1697]{Christopher~T.~Garling}
\affiliation{
Space Telescope Science Institute,
3700 San Martin Dr, Baltimore, MD 21218, USA
}
\affiliation{%
Department of Astronomy, 
University of Virginia,
530 McCormick Rd, Charlottesville, VA 22904, USA}
\affiliation{The NSF-Simons AI Institute for Cosmic Origins, USA}
\email{Email:\ cgarling@stsci.edu}

\author[0000-0002-7393-3595]{Nathan~R.~Sandford}
\affiliation{Department of Astronomy and Astrophysics, University of Toronto, 50 St. George Street, Toronto ON, M5S 3H4, Canada}
\affiliation{Department of Astronomy and Astrophysics, University of California, Santa Cruz, 1156 High Street, Santa Cruz, CA 95064, USA}
\email{Email:\ nasandfo@ucsc.edu}

\author[0000-0003-2806-1414]{Lina~Necib}
\affiliation{Department of Physics and MIT Kavli Institute for Astrophysics and Space Research, \\
Massachusetts Institute of Technology,
77 Massachusetts Avenue, Cambridge, MA 02139, USA}
\affiliation{The NSF AI Institute for Artificial Intelligence and Fundamental Interactions, \\
Massachusetts Institute of Technology,
77 Massachusetts Avenue, Cambridge, MA 02139, USA}
\email{Email:\ lnecib@mit.edu}

\author[0009-0002-1233-2013]{Niusha~Ahvazi}
\altaffiliation{Galaxy Evolution and COsmology (GECO) Fellow}
\affiliation{%
Department of Astronomy, 
University of Virginia,
530 McCormick Rd, Charlottesville, VA 22904, USA}
\affiliation{The NSF-Simons AI Institute for Cosmic Origins, USA}
\email{Email:\ dkk9en@virginia.edu}

\author[0009-0000-0733-2479]{Andr\'es~Almeida}
\affiliation{%
Department of Astronomy,
University of Virginia,
530 McCormick Rd, Charlottesville, VA 22904, USA}
\email{Email:\ tac6na@virginia.edu}

\author[0000-0001-9649-8103]{Kaia~R.~Atzberger}
\affiliation{%
Department of Astronomy, 
University of Virginia,
530 McCormick Rd, Charlottesville, VA 22904, USA}
\email{Email:\ yzb4en@virginia.edu}

\author[0009-0002-8461-6111]{Yanbo~Pan}
\affiliation{%
Department of Astronomy, 
University of Virginia,
530 McCormick Rd, Charlottesville, VA 22904, USA}
\email{Email:\ war8sk@virginia.edu}

\author[0000-0003-1634-4644]{Jack~T.~Warfield}
\affiliation{%
Department of Astronomy, 
University of California, Berkeley,
Campbell Hall, Berkeley, CA 94720, USA}
\affiliation{%
Department of Astronomy, 
University of Virginia,
530 McCormick Rd, Charlottesville, VA 22904, USA}
\email{Email:\ warfield@berkeley.edu}

\begin{abstract}

Ultra-faint dwarf galaxies (UFDs) in the Milky Way are ideal probes of low-mass galaxy formation and dark matter because they are highly dark-matter dominated.
In the lowest-mass systems, however, unresolved binary orbital motion complicates dynamical mass estimates by inflating measured velocity dispersions and biasing the interpretation of the inferred masses.
We develop a flexible forward-modeling pipeline, the Binary Observation Simulator, to generate mock multi-epoch radial-velocity observations of binary populations in dwarf galaxies and to quantify how binary-induced biases depend on both galaxy properties and observing strategy.
We generate mock samples spanning true velocity dispersion of $\sigma_{\rm true}\sim0.75$--$3.5$ \kmsec, binary fractions of $0.1$--$0.9$, and multi-epoch baselines of up to $10$ yr.
We find that, while multi-epoch monitoring reduces the binary-induced bias, residual contamination remains significant for low-mass halos: even a $10$-yr baseline can leave $\sim10$--$120\%$ relative bias for systems with $\sigma_{\rm true}\lesssim1$\,\kmsec.
We also note that small sample sizes ($\lesssim20$) introduce substantial stochastic scatter in the recovered dispersions, sometimes masking the expected improvement from longer observational baselines.
Applying the framework to the real $\sim17$-yr \bootes~I observing record reproduces the literature velocity dispersions, yields an empirical correction conditioned on that cadence, and shows that redistributing the same observations onto half the number of stars would have removed roughly half of the residual binary bias.
These results emphasize that robust dispersion estimates for the faintest dwarfs require forward modeling of both binary populations and survey cadence.

\end{abstract}


\section{Introduction} 

Ultra-faint dwarf galaxies (UFDs) provide one of the most powerful observational windows into galaxy formation and the nature of dark matter on small scales ($<1$kpc) (see, e.g., \citealt{kroupa10,bullock17,perivolaropoulos22}).
They are valuable probes for studying the least massive scale of galaxy formation with $M_{\rm star}\lesssim 10^{5}$\,\msun and $M_{\rm halo}\lesssim 10^{7-9}$\,\msun \citep{simon07,strigari18, simon19, nadler20, ahvazi24}. 
Furthermore, the mass function of dwarf galaxies (e.g., \citealt{jethwa18,kim18,nadler19,mau22}) and the internal structure of their dark matter halos (e.g., \citealt{calabrese16,errani18,bozek19,sales22,silverman23,ahvazi25}) are central observables for testing the theoretical expectations from dark matter models for low-mass halos.

In the Local Group, access to resolved stellar populations allow spectroscopy measurements that are inaccessible for more distant systems, enabling kinematic constraints based on line-of-sight velocities \citep[e.g.][]{cn+21,guerra23}. 
The gas-free and pressure-supported environment makes UFDs comparatively clean laboratories for using stellar kinematics to probe the underlying gravitational potential \citep{brown14, simon19, battaglia22a}. 
For example, under assumptions of dynamical equilibrium and approximate spherical symmetry, the enclosed dynamical mass within the half-light radius can be relatively robustly estimated from the velocity dispersion \citep[e.g.,][]{wolf10}. 
This approach is particularly important for the faintest systems, where the available data are often limited to projected positions and radial velocities for tens to hundreds of stars. 

While the velocity dispersion serves as a practical first-order mass estimate in the observational regime, this estimator faces observational and astrophysical systematics that can significantly bias the results.
This is especially true for UFDs, whose intrinsic velocity dispersions can be only a few \kmsec \ or less. 
In this low-dispersion regime, foreground contamination, small-number statistics, non-equilibrium dynamics, and unresolved binary orbital motion can each contribute a velocity scale comparable to the signal being measured \citep{simon19}. 
Consequently, robustly interpreting the low-mass end of the dwarf galaxy population requires not only measuring velocity dispersions, but also quantifying the extent to which those dispersions may be inflated or biased.

In particular, unresolved binary stars are an important source of uncertainty for UFD kinematic studies. 
For dynamically cold systems, binary orbital velocities can be comparable to or larger than the intrinsic galaxy dispersion, causing the observed velocity distribution to broaden even when all stars are genuine members (see e.g., \citealt{spencer17,pianta22}). 
In extreme cases, ultra-faint stellar systems with globular-cluster-like intrinsic dispersions of only $\sim0.2$\,\kmsec \ and an unidentified binary population can produce apparent dispersions $\sim4.5$\,\kmsec\ \citep{mcconnachie10}.
These binary induced biases can be addressed with either single or multi-epoch observations of the stars in the systems.

With single epoch measurements, one can principally perform joint Bayesian inferences of the intrinsic velocity dispersion and the binary fraction simultaneously for a given system (see e.g., \citealt{minor10, arroyopolonio26}).
While the inference does not identify any particular star as a potential binary, it models explicitly the binary-induced biases in the velocity dispersion of a sample of stars. 
It has further been shown that the intrinsic dispersion inferred via such binary-aware likelihood functions are relatively robust against assumptions about the underlying binary population properties (mainly period) both for systems with star sample sizes $>100$ \citep{minor10} and sample sizes $<100$ \citep{arroyopolonio26}. 
However, in cases of very small sample sizes ($\lesssim30$) for UFDs, the stochastic effect, combined with even a low binary fraction ($<0.3$), can result in measured dispersions inflated by over factor of a few for systems with intrinsic dispersions $\sim1$\,\kmsec \citep{wang23,arroyopolonio26}.
While rare, joint inferences based on single epoch measurements become insufficient to recover the intrinsic dispersion in these extreme cases. 

Multi-epoch spectroscopy, when available, is therefore preferred for directly identifying and removing velocity-variable stars as potential binaries, reducing binary-induced biases in the system velocity dispersion. 
Most recent studies (see \citealt{simon19} and \citealt{pace24} for a summary of UFD kinematic studies) utilize the multi-epoch measurements by performing a binary-exclusion procedure before inferring the velocity dispersion at each epoch.
Even two epochs of observations spaced at a 1-year baseline are shown to improve the robustness of velocity dispersion measurements significantly, without the need to jointly model the binary fraction at each epoch \citep{minor10,wang23,arroyopolonio26}.
Furthermore, longer baseline multi-epoch monitoring of UFDs (e.g., \bootes\,I; \citealt{sandford26}) has found that the inferred velocity dispersion continues to decrease as more binaries are identified and removed at each additional epoch.
Such a trend highlights the importance of long baselines, in addition to measurement precision. 

Yet, the necessary baseline length is often unclear for different systems and instrument capabilities. 
In other words, is there a point at which we can confidently argue that no longer baseline observation is required?
A naive hard finish line may be when the multi-epoch baseline is longer than the typical period of binaries that have \textit{velocity amplitudes close to the uncertainty floor of velocity measurements}, at which point one may argue that any further observation becomes futile as we can no longer identify binaries.
It is a reasonable argument until one starts to consider cases of highly eccentric and/or extreme mass ratio binary orbits, at which point the definition of ``typical period of binaries with a certain velocity amplitude'' becomes blurry \citep{minor10}.
Further complicating the picture, the ever changing landscape of observing instrument capability means that the spectral and spatial resolution changes with time.
The former directly determines the velocity measurement precision, while the latter indirectly affects our ability to constrain the binary contribution through visual (resolved) \citep[e.g.,][]{shariat25} and spectroscopic (unresolved) binaries \citep[e.g.,][]{filion21,gration25}.
Lastly, while a finite-baseline survey can identify systems whose measured velocities vary by more than some threshold, it does not guarantee removal of binaries whose time-averaged measured velocity over the observed epochs remains offset from the binary center-of-mass velocity by a comparable amount, as highlighted in recent survey forecasts \citep[e.g., Section~3.6.2 of][]{via26}. 
Only in the limit of infinitely long and well-sampled baselines does the observed mean velocity converge to the systemic velocity for each binary. 
Quantifying these residual effects is then conditioned on the existing multi-epoch observation baseline. 
Thus, these questions must be addressed on a galaxy-by-galaxy basis.
For example, \citet{martinez11} and \citet{minor19} utilize an inference framework that self-consistently models multi-epoch observations simultaneously for Segue\,I and Reticulum\,II, respectively.
Meanwhile, the majority of the UFD velocity dispersions are still inferred from a binary-free likelihood function, after potential binaries are removed.

Taken together, the challenges arise from the heterogeneous nature on the observational end, and the high dimensionality of free parameters on the theory end.
Stars of the same galaxy are typically observed across varying baselines, with distinct instruments and data reduction systematics.
The resulting measurements are difficult to model with a simple likelihood function with some fixed observational baseline or measurement uncertainty. 
Moreover, binary rejection itself can alter the effective member sample, coupling the inferred dispersion to the details of the observing strategy and the adopted binary-cleaning procedure. 
Up to very recently, there was not even a centralized database for existing measurements in the literature, although efforts are being made \citep[see e.g.,][]{geha26}.

In light of these challenges, it is thus crucial to create a flexible inference framework that can efficiently sample multi-epoch velocity measurements of stars, conditioned on both intrinsic galaxy properties, such as intrinsic dispersion and binary fraction, and observing strategy, such as epoch and instrument uncertainties.
This study serves as a first step by constructing the flexible forward modeling component of such a framework (Section~\ref{sec:methods}), enabling future work to perform inference with observational data as the forward modeling efficiency improves.
We utilize measurements generated by the model (Section~\ref{sec:mock_sample}), combined with mock velocity dispersion analysis (Section~\ref{sec:mock_analysis}), to first gain insights on the binary-inflated dispersion as a function of galaxy properties and observing conditions (Section~\ref{sec:disc_mock}).
We then show that the model can reproduce results from literature UFD velocity dispersion studies with minimal finetuning to the model parameters (Section~\ref{sec:case_bootes1}).
Furthermore, the model is useful for projecting the expected yield of alternative observation strategies in identifying and reducing biases due to binaries (Section~\ref{sec:strategy_pred}).
Lastly, we show that the model can be used to provide an empirical correction for derived velocity dispersions for UFDs, conditioned on arbitrary model parameters, including observation conditions such as baseline and instrument precision (Section~\ref{sec:empirical_corr}).
We discuss future steps and conclude in Sections~\ref{sec:limit} and~\ref{sec:conclusion}, respectively.

\section{Methods}
\label{sec:methods}

We describe the structure of our mock observation pipeline, the Binary Observation Simulator (BOS), in this section.
Specifically, the BOS implements a forward-modeling workflow that generates realistic mock observations of single and binary star systems in dwarf galaxies. 

The global properties of the dwarf galaxy, including dark matter halo mass, stellar mass, and structural parameters such as half-light radius, are set first. 
These properties can be sampled from empirical scaling relations, specified through custom galaxy models (e.g., \citealt{burkert95,navarro97}), or set manually by the user. 
The stellar density profile is then computed, typically using analytical models such as the Plummer profile \citep{Plummer:1911}, which determine the spatial distribution of stars and enable the calculation of velocity dispersion profiles through the Jeans equation.
Three-dimensional positions for all stars are sampled from the stellar density profile.
Velocities are similarly assigned by sampling from velocity dispersion profiles, accounting for radial and tangential velocity components and optional velocity anisotropy.

At the same time, individual stellar masses are sampled from an initial mass function (IMF; e.g., \citealt{kroupa10}). 
A fraction of the stellar population is then randomly assigned to be the primary stars in binary systems based on a specified binary fraction.
For each binary system, orbital parameters, including orbital periods ($P$), mass ratios ($q$), and eccentricities ($e$), are sampled from population models such as \citet{dupuennoy91,moe17}, implemented by \citet{gration25b} in \textsc{dyad}. 
In the case of the \citet{moe17} model, the binary property distributions are conditioned on the stellar mass of the randomly selected primary stars. 
Once the mass ratios are determined, the best possible secondary companions are chosen from the remaining single stars and paired with each primary.
Specifically, the pairing scheme follows a greedy search for the best plausible secondary given the mass ratio.
This approach is adopted to ensure maximum consistency between the IMF and the mass ratio distributions.

After binary pairing, the primary positions and velocities are assigned to each binary system as the center-of-mass 6D kinematics.
The orbital phase, inclination, and argument of periastron are randomly sampled for each binary.
For both the single and binary systems, stellar ages and metallicities are sampled from user-specified star formation history and metallicity distribution models. 
Stellar evolutionary properties (magnitudes, evolutionary stages) are then computed using isochrone interpolation based on the stars' ages, metallicities, and masses, using the \textsc{StellarTracks.jl} interpolator tool built by \citet{garling25}.

After generating the intrinsic properties of the single and binary systems in a dwarf galaxy, the simulator generates mock observations.
Multi-epoch observations are generated by computing the line-of-sight velocities of all stars at some given observation epoch. 
For binary systems, the orbital motion is evolved forward in time and added to the systemic velocity of the system. 
Realistic observational errors are then added, following magnitude-dependent error models that account for the decreasing signal-to-noise ratio for fainter stars. 
Additional observational effects can be included, such as systematic instrument offsets between epochs and random incomplete sample coverage.
The final output consists of multi-epoch radial velocity measurements with associated uncertainties, observed magnitudes, and metadata describing the observational conditions and binary system properties.

This modular workflow allows users to customize or bypass any step, enabling flexible exploration of different physical models and observational scenarios.
Specifically, an alternative mode is provided for cases where the stellar sample can be set to mirror observed samples in real UFDs.
In this mode, the user supplies a stellar catalog directly: only initial masses are required, while ages, metallicities, three-dimensional positions, systemic velocities, and photometry may each be either supplied or delegated to an analytical model described above.
To keep the mock observed sample size constant, binaries are then introduced by \emph{spawning} unobserved companions rather than by pairing within the supplied catalog.
Each catalog star is independently flagged as a binary primary with probability equal to the assumed binary fraction, and a companion of mass $q\,M_{\rm primary}$ is appended to the catalog, sharing the primary's age, metallicity, and center-of-mass phase-space coordinates.
For mock observations, the alternative mode accepts a per-star observing schedule specifying, for each star, the epochs at which it was observed together with the corresponding velocity uncertainty and instrument offset.
The mock analysis can subsequently be restricted to exactly the measurements a real campaign obtained.

The two modes are complementary.
The default pairing conserves the input mass function exactly, since companions are drawn from stars the IMF already produced, and is the appropriate choice when the goal is a self-consistent population with controlled global properties, as in the mock grid of Section~\ref{sec:res}.
The alternative spawning does not conserve the input mass function, but it leaves the supplied catalog untouched, which makes it the natural interface for confronting the simulator with a specific observed system, with a real stellar member catalog, observation strategy, and analysis pipeline.
We use the default pairing mode throughout Sections~\ref{sec:res} and~\ref{sec:disc_mock} for studying general binary residual in three fiducial test cases, and switch to the alternative mode in Section~\ref{sec:link_obs} for comparison with literature.

\section{Mock velocity dispersion for fiducial cases}
\label{sec:res}

\subsection{Mock sample generation}
\label{sec:mock_sample}

In light of recent discoveries of ultra-faint compact satellites \citep[see e.g.,][]{cerny26}, three sets of mock samples are produced with parameter combinations covering three characteristic halo masses at $M_{\rm 200} \sim10^{4.5},10^{5.5},10^{6.5}$\,\msun. 
The dark matter halos are all assumed to be an NFW profile with concentration consistent with the halo mass-concentration relation from \citet{dutton14}. 
We choose this mass range to probe systems with expected intrinsic dispersion $\lesssim4.5$\,\kmsec, where binary orbital motion can plausibly contribute a large fraction of the measured dispersion, informed by results from \citet{mcconnachie10}.

For the corresponding stellar profiles, we arbitrarily assume Plummer profiles with stellar mass and scale radius parameters such that $M/L\sim100$ and the intrinsic dispersion of a binary free sample yield $\sim0.75$, $\sim1.5$, and $\sim3.5$\,\kmsec, respectively for the three halo mass cases.
While it is more ideal to initialize the dark matter and stellar profiles with N-body simulation to ensure dynamical equilibrium, our results are qualitatively unchanged as long as the systems are dark matter dominated.
All mock sample sizes are set to $10000$ stars for computational efficiency.
We apply additional down-sampling in the mock analysis stages through sparse observational completeness for more realistic sample sizes at each observing epoch.

\begin{figure}
    \centering
    \includegraphics[width=0.95\linewidth]{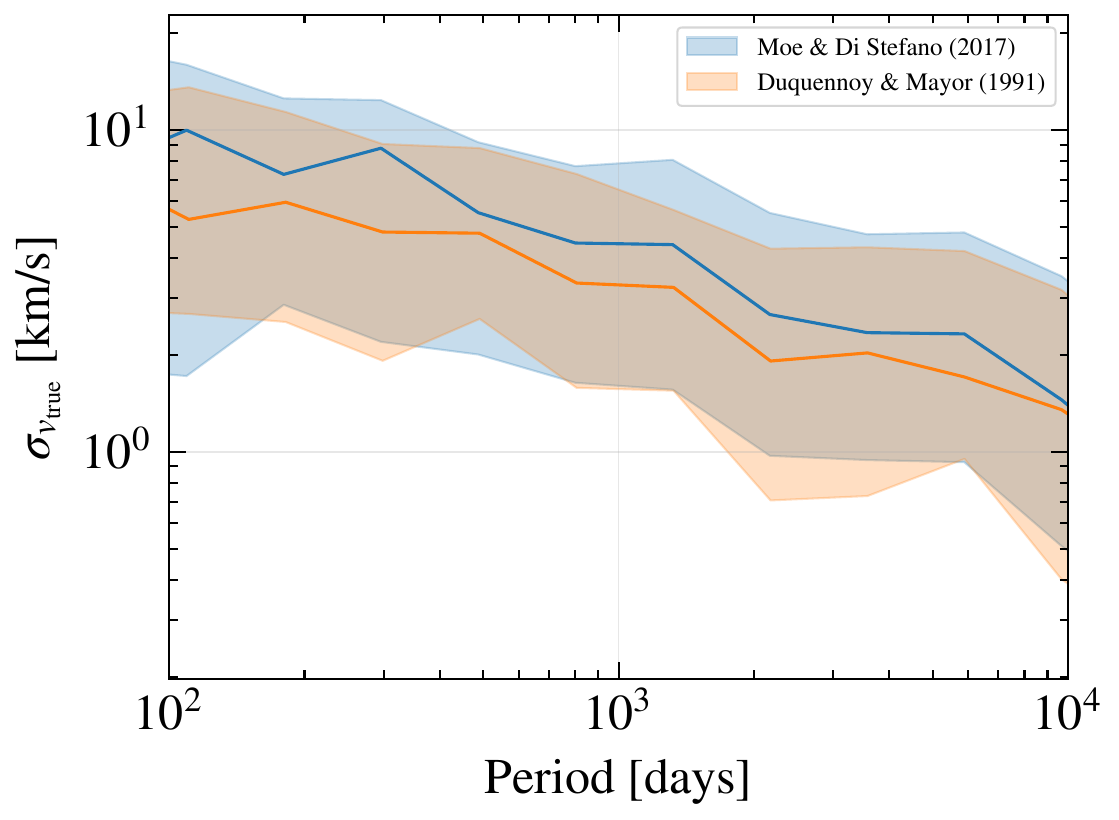}
    \caption{
    Comparison between the scatter induced by binary motion in the line-of-sight velocities of binary systems sampled from \citet{dupuennoy91} (yellow) and \citet{moe17} (blue) models.
    }
    \label{fig:bin_model_compare}
\end{figure}

For individual stars, initial stellar masses are sampled from a Kroupa IMF \citep{Kroupa:2001}.
Stellar ages are drawn from a uniform star formation model with ages between $13.5\pm0.1$\,Gyr. 
Metallicities are assumed to follow a Gaussian MDF with mean $\rm{[Fe/H]}=-2.0$, $\sigma = 0.5$\,dex, and minimum $\rm{[Fe/H]} = -4.0$. 
Stellar evolutionary properties are interpolated based on the MIST isochrone set with Gaia EDR3 photometric bands.
These model configurations are selected to represent ancient, quenched, and metal-poor stellar populations, typical for UFDs considered in this study. 
Binary system properties are generated from the \citet{moe17} model.
We find the results of the analysis to be relatively insensitive to the specific choice of the binary population models given the mock observational conditions described below.
As shown in Figure~\ref{fig:bin_model_compare}, the binary-induced scatter in the line-of-sight velocities from different epochs for any given binary star follows qualitatively the same trend for both the \citet{dupuennoy91} and \citet{moe17} binary population models.
For each halo mass case, we sample five binary fractions ranging from $0.1$ to $0.9$ in steps of $0.2$.

For observational parameters, all galaxies are assumed to be at $20$\,kpc away from the observer for computing the apparent magnitude.
We adopt magnitude-dependent errors for both the photometry and the radial velocity measurements. 
Specifically, radial velocity errors follow a power-law of $\sigma = 0.1\cdot 10^{0.4(m_G-18)}$\,\kmsec, with a floor of $0.1$\,\kmsec, qualitatively matching the next generation spectroscopic survey \citep{via26}. 
For the majority of this study, the line-of-sight velocity measurement uncertainties range from $0.1$ to $\sim 2$\,\kmsec \ from $G<20$ to $G<21$.
Similarly, we assume a power-law for the photometric magnitude errors of $\sigma = 0.005\cdot 10^{0.4(m_G-20)}$\,mag, with a floor of $0.005$\,mag. 
Both error models are configured for all epochs of observations.
Observations are taken at four epochs: 0, 365, 1825, and 3650 days (0, 1 year, 5 years, and 10 years). 

We perform 100 iterations per mock sample for any combination of the three halo masses and the five binary settings to focus on the medium trend across the 100 iterations for the majority of this study, avoiding stochastic outliers from biasing our interpretation.
A total of $1500$ mock samples are thus generated for this study.
We discuss the stochasticity from individual iterations and their impact on interpretation of the inferred velocity dispersion in Section~\ref{sec:disc_stat}.

\subsection{Mock analysis}
\label{sec:mock_analysis}

For each iteration of a mock sample with some given intrinsic mass scale and binary fraction, we conduct a mock analysis to identify and remove stars in binary systems based on multi-epoch data, subsequently deriving the velocity dispersions.

For each realization of a given mock analysis, an initial observable sample is first selected from the parent $n=10000$ mock sample based on the observed $G$-band magnitudes, to examine the dependence of the analysis results on the sample size.
Practically, we consider two cases of observing depth, $G<20$ and $G<21$, corresponding to (1) stars on the giant branches and (2) stars down to the MSTO, respectively.
For the assumed IMF and stellar evolution model, $\sim30$ and $\sim100$ stars from the initial sample meet the two magnitude cuts at $G<20$ and $G<21$, respectively. 
Thus, we perform the following mock analysis on these two sample sizes to mimic two cases of dwarf galaxy dynamics studies: (1) smaller sample sizes but at high per star velocity precision and (2) larger sample sizes but with worse per star velocity precision.

We additionally test the effect of sparse coverage completeness in the multi-epoch observations.
Due to limited observational resources and constraints, it is often the case that not all observable stars in a dwarf galaxy will be observed at every epoch.
In most cases, as membership selection evolves, new targets are identified and added to the sample.
Such incomplete multi-epoch observations result in both reduced sample size and limited ability to identify binaries.
Thus, it is also pertinent to study the effect of incomplete visits across all available epochs for a given mock sample.

In practice, we approximate this effect by randomly selecting a fraction of the observable stars at each epoch to be observed.
We consider two cases with the random observing fraction of $50\%$ and $80\%$ of all the observable stars.
The resulting mock samples would thus have stars with the number of observations spanning from a minimum zero to a maximum equal to the number of epochs.
For any given epoch, we perform the mock analysis following the procedure detailed below, utilizing all available baselines to exclude binaries from the sample and fit for the velocity dispersion.

The mock analysis cleans the sample cumulatively at each observing epoch, using every measurement obtained up to and including that epoch.
Two criteria are applied in sequence. 
First, stars whose repeat measurements are inconsistent with a constant velocity are flagged as binaries and removed.
For each star, we compute the $\chi^2$ of its measurements about their inverse-variance weighted mean and reject those with $p<0.05$.
The surviving sample is then iteratively sigma-clipped. 
We fit the systemic velocity $v_0$ and intrinsic dispersion $\sigma$ of the retained stars, reject stars deviating from $v_0$ by more than $3\sqrt{\sigma^2+v_{{\rm err},i}^2}$, and repeat until the retained sample is stable.
Because the clipping threshold scales with the \emph{fitted} dispersion rather than an assumed one, it adapts to the system being analyzed.
Alternatively, the analysis can instead apply a single $5\sigma$ membership window constructed from an assumed intrinsic dispersion to remove the outliers.
In this case, the results are sensitive to the choice of $\sigma$: while a wide window helps retain the true high-amplitude tail of the velocity dispersion, a narrow one removes tight binaries and potential foreground contaminants more effectively.
We therefore adopt the data-driven clip described here, which introduces no such free parameter.
We note that the choice of cleaning scheme is itself a source of systematic uncertainty, and one that can alter the observing-strategy conclusions drawn from the mocks rather than merely rescaling them; we quantify this in Appendix~\ref{sec:app_cleaning}.

Although every star in the mock sample is by construction a member of the galaxy\footnote{Our model currently does not include a Milky Way foreground model, we assume a \Gaia\ proper motion/parallax cleaning and/or CMD selections remove most of the MW foreground before the velocity is considered.}, $\sim7$--$17\%$ of the sample observed at the first epoch (across the three fiducial cases and both depths) is nonetheless removed by the velocity clip. 
These are predominantly stars in close binary systems whose orbital phase at the time of observation produces a large excursion from the systemic velocity.
In a real analysis such stars would be rejected as non-members, or flagged as binaries once a second epoch was obtained.
We therefore treat their removal as part of the cleaning rather than as a loss of genuine members.

After the cleaning at each epoch, the velocity dispersion is fit using \textsc{emcee}. 
The systemic velocity ($v_0$) and velocity dispersion ($\sigma$) are fitted simultaneously via the likelihood function,
\begin{equation}
    \log{\mathcal{L}} = \sum^{N}_{i=1} \left( \log{\frac{1}{\sqrt{2\pi(\sigma^2+v_{\rm{err},i}^2)}}} \right) - \sum^{N}_{i=1} \left( \frac{(v_{r,i}-v_0)^2}{2(\sigma^2+v_{\rm{err},i}^2)} \right),
\end{equation}
where $v_{r,i}$ and $v_{\rm{err},i}$ are the measured line-of-sight velocity and associate uncertainty of star $i$.

In addition to the mock dispersion analysis at each epoch, we perform a velocity dispersion fit using the true line of sight velocities of all observable stars, which yields $\sim0.75$, $\sim1.5$, and $\sim3.5$\,\kmsec, respectively for the three test cases. 
The result serves as a proxy for the true velocity dispersion for comparison with those from the mock analysis and we examine the relative bias (defined in this study as $\sigma_{\rm fitted}/\sigma_{\rm true}$) for the majority of the discussion to highlight the fact that low mass halos are more susceptible to binary contamination.
These relative biases assess how multi-epoch observations for removing non-member and binary systems reduces the bias in the measured velocity dispersion from unaccounted for binary orbital motion.

Following the mock analyses, each mock sample produces its own relative bias measurement as a function of epoch. 
As discussed in Section~\ref{sec:mock_sample}, there are 100 random iterations of a mock sample with any given intrinsic halo mass and binary fraction setting.
We thus combine and report the results from the corresponding 100 mock analyses.
The medians of the 100 measured relative biases and associated uncertainties are taken as the nominal relative bias and uncertainty at each epoch.

\begin{figure}
    \centering
    \includegraphics[width=1.0\linewidth]{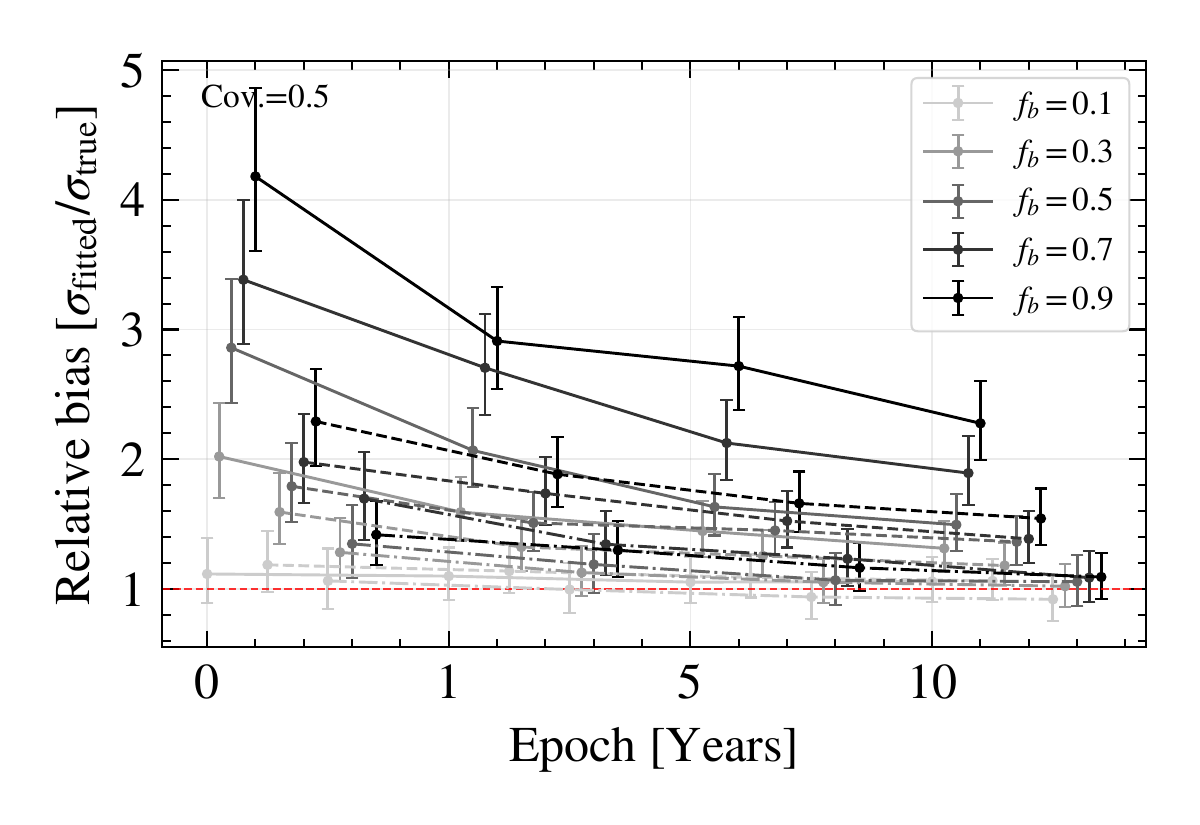}
    \caption{
    The relative velocity dispersion bias as a function of observing epochs for three halos with 
    $\sigma_{\rm true} = 0.75$, $\sigma_{\rm true} = 1.5$, and $\sigma_{\rm true} = 3.5$\,\kmsec\ in solid, dashed, dash-dotted lines, respectively. 
    The different shades of lines represent analysis results of mock samples with input binary fraction from $0.1$ (lightest shade) to $0.9$ (darkest shade).
    The observational depth is set to $G<20$, while the observing fraction (Cov) is $50\%$.
    The relative velocity dispersion biases and associated error bars at each epoch are computed from the 100 iterations of mock analyses as described in the text.
    The dashed line at $\sigma_{\rm fitted}/\sigma_{\rm true}=1$ corresponds to the case with no relative bias.
    Data points are offset horizontally to the right from the mock observation epochs at $0$, $1$, $5$, and $10$ years for clarity.
    }
    \label{fig:mock_analysis}
\end{figure}

We present the results of the mock analysis on the three halo mass cases with varying binary fraction in Figures~\ref{fig:mock_analysis}.
We assume an observational depth of $G<20$ and observing fraction of $50\%$ for all cases, resulting in $\sim15$ out of the $\sim30$ observable stars observed at each epoch and $\sim60\%$ of the stars having at least two observations.
These conditions are optimistic compared to typical observational samples available for UFDs in the Milky Way, but they are chosen to be uniform for a straightforward comparison and interpretation of the different factors at play.
We stress that these conditions should and can be adjusted to match conditions available to specific real UFDs before comparison with the literature (as shown in Section~\ref{sec:case_bootes1}).

\section{Relative biases in the fiducial cases}
\label{sec:disc_mock}

In this section, we examine the general test cases described in the last section and discuss the dependence of the relative bias ($\sigma_{\rm fitted}/\sigma_{\rm true}$) on the galaxy properties and observation strategies.
We specifically examine the relative bias as a function of multi-epoch observation baselines, conditioned on the galaxy mass scales and the binary fractions in the system.
We then briefly discuss the effect of small sample sizes and stochasticity on the derived dispersions as a result of limited observing resources.
We lastly focus on the expected ability to reduce the relative biases by applying different observing strategies. 

\subsection{Dependence on galaxy mass and binary fraction}
\label{sec:dep_galprop}

Across all mock realizations, we find that unresolved binary orbital motion systematically inflates the inferred line-of-sight velocity dispersion when only single-epoch or short-baseline measurements are available, even for dwarf galaxies with binary fractions as low as $\sim0.1$. 
Specifically, at the first epoch, the recovered velocity dispersion is consistently biased high relative to the intrinsic dispersion traced by single stars alone.
For typical UFD-like systems, this overestimation ranges from $\sim10\%$ at $f_{\rm b}=0.1$ to a factor of $\sim4$ at $f_{\rm b}=0.9$, corresponding to an absolute bias of up to $\sim2.2$\,\kmsec\ for the least massive system with an intrinsic dispersion of $\sim0.75$\,\kmsec.
These findings are expected: in the absence of multi-epoch information, velocity dispersion estimates for low-dispersion systems should be interpreted as upper limits rather than unbiased measurements, even when standard sigma-clipping and membership cuts are applied.

Introducing multi-epoch radial-velocity measurements mitigates the bias from binary systems, even in cases when only $\sim50\%$ of the stars are revisited in the second epoch and the remaining $50\%$ are new members observed for the first time, as shown in Figure~\ref{fig:mock_analysis}.
However, a residual bias remains for systems with binary fraction from $\gtrsim0.1$ to $\gtrsim0.5$, for galaxies with intrinsic dispersions from $0.75$ to $3.5$\,\kmsec, due to (1) close binaries that are only observed once and (2) wider binaries that have velocity not varying significantly at 1-year baseline.
For the more massive cases with intrinsic dispersion of $1.5$ ($3.5$)\,\kmsec, this means that multi-epoch monitoring on timescales of order one year is often sufficient to recover dispersions that are statistically consistent within 2\,$\sigma$ with the single-star population, for binary fraction up to $\sim0.3$($\sim0.5$).
By contrast, for the lowest-mass systems with intrinsic dispersions below $\sim1.0$\,\kmsec, short 1-year baseline monitoring is not sufficient to recover the intrinsic dispersion, with overestimated dispersion of order $\gtrsim50\%$ even at binary fraction $\sim0.3$. 
Even with a five-year baseline, residual biases of $\sim50\%$ remain common for binary fraction $\gtrsim0.3$.

Extending the baseline to $\sim$10 years further suppresses the contribution from long-period binaries, fully eliminating the dispersion overestimate for the most massive halo case.
Yet, the relative bias remains significant at $\sim10$--$120\%$ for the least massive halo case.
This remaining bias is driven by the fact that the surviving sample still contains binaries that are not identified as a result of incomplete coverage at every epoch.
These leftover binary contaminants may contribute up to only $\sim0.25$\,\kmsec\ to the absolute velocity dispersion.
Yet, while $\sim0.25$\,\kmsec\ may be negligible for systems with intrinsic dispersions of $\gtrsim1.5$\,\kmsec, the same bias translates into a fractional overestimate of order $\sim30\%$ for systems with intrinsic dispersions near $\sim0.75$\,\kmsec. 
This mass-dependent amplification highlights why binary contamination is most problematic for the faintest and dynamically coldest dwarf galaxies.

Across all three mass scales explored, we find that the binary fraction itself is a sub-dominant driver of the relative bias, especially at values $\gtrsim0.5$, compared to the galaxy mass. 
The binary fraction affects the relative bias the most at low intrinsic velocity dispersion cases.
For high binary fraction $\gtrsim0.5$ in the most massive halo, variations in binary fraction from $0.7$ to $0.9$ poses minimum changes to the relative dispersion bias.
This insensitivity of the relative bias to the binary fraction hints that it is possible to empirically estimate the binary-induced bias at a given epoch without prior knowledge of the true binary fraction, which we discuss in detail in Section~\ref{sec:empirical_corr}. 

\subsection{Dependence on observing strategy}
\label{sec:sparse_visit}

\begin{figure*}
    \centering
    \includegraphics[width=0.95\linewidth]{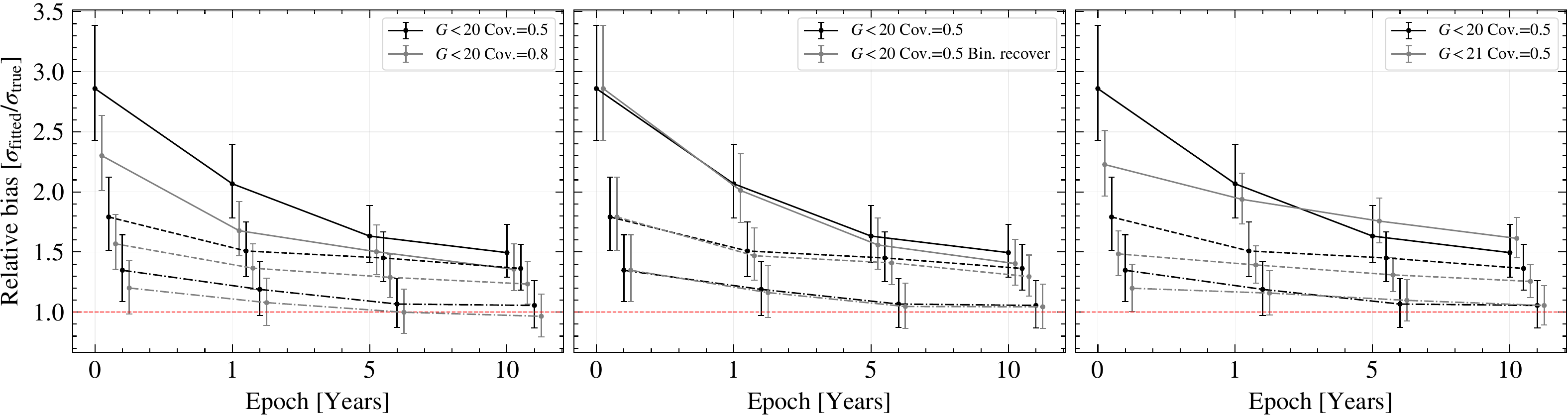} 
    \caption{
    Comparisons examining the effect on the relative bias in velocity dispersions as a function of baseline due to different observing strategies.
    We adopt the mock observation+analysis scenario from Figure~\ref{fig:mock_analysis}, with $G<20$, coverage fraction of $50\%$, and all identified binaries removed from the sample, as the baseline.
    We examine three halo mass scales (same line style as in Figure~\ref{fig:mock_analysis}) with an assumed binary fraction of $0.5$.
    The left panel shows the improvement in relative bias from increasing sample coverage at every epoch from $50\%$ (black) to $80\%$ (gray).
    The middle panel shows the improvement from recovering the systemic velocities of all identified binaries and incorporating them back in the dynamical analysis.
    The right panel compares observational depths at fixed coverage, which can have both positive and negative effects depending on whether sample size or velocity precision is the limiting factor.
    Data points are offset horizontally to the right from the mock observation epochs at $0$, $1$, $5$, and $10$ for clarity.
    }
    \label{fig:improv}
\end{figure*}

Moving on from the intrinsic properties of the galaxies, we now examine the effect of observational strategy choices in the relative biases as a function of multi-epoch observations: from varying observational depth to fraction of stars visited at each epoch.
With limited observational resources, it is often a question whether one should devote more effort into securing fainter stars for a larger sample size with fewer observations for each target, focus on securing more complete multi-epoch observations for a smaller sample of bright stars, or even dedicate observing time to recovering identified binaries' systemic velocities.
Specifically, we consider three cases of possible observational strategies to reduce the relative bias from our baseline scenario: (1) increasing the fraction of stars revisited at each epoch from $50\%$ to $80\%$, (2) recovering the systemic velocities of all identified binaries, and (3) acquiring deeper observation for stars from $G<20$ to $G<21$, as shown in Figure~\ref{fig:improv}.
For simplicity, we focus on a binary fraction of $0.5$ for all three halo mass cases in the following discussion.

We find generally good improvement in the relative bias for strategies (1) and (2), with the former outperforming the latter.
For the least massive halo, the relative bias reduced from $\sim50\%$ to $35\%$ at the final 10-year epoch with a $80\%$ observing completeness down to $G<20$.
The same change brings the intermediate halo from $\sim30\%$ to $\sim20\%$.
Recovering all the binaries has a similar, but less expedient positive effect, reducing the relative bias of the least massive halo to $\sim45\%$ only at the last epoch.
Under the present cleaning scheme, only the stars flagged as velocity variables by the $\chi^2$ test are recovered, so the improvement only becomes apparent when more binaries are revisited. 
We assumed that all identified binaries at each epoch can be followed up and solved for their systemic velocities.
While this is theoretically plausible for any binaries that show significant velocity variation at these epochs, realistic observations are typically subject to limits such as telescope availability and thus the improvement should be treated as an optimistic estimate.
We note that stars rejected by the sigma-clipping on the basis of a single discrepant measurement are not recovered, since in real data such a star cannot be confidently distinguished from a non-member.

The effect of observational depth is more subtle, and depends on both the halo mass and the available baseline.
At the first epoch, the deeper $G<21$ sample is the less biased of the two for all three halos because the velocity clip is the only cleaning available at a single epoch and it is more stable on a sample of $\sim100$ stars than on one of $\sim30$.
As the baseline lengthens, this advantage erodes.
The larger velocity uncertainties at fainter magnitudes, which grow from $\sim0.5$\,\kmsec\ at $G=20$ to $\sim2$\,\kmsec\ at $G=21$ (see Section~\ref{sec:mock_sample}), progressively weaken the ability of the $\chi^2$ test to identify binaries, so a greater fraction of long-period binaries (velocity variations $\sim1$\,\kmsec) survives in the deeper sample.
For the two more massive halos the deeper sample nonetheless remains the less biased one at every epoch, since their intrinsic dispersions are large enough that the added uncertainty is not the limiting factor.
For the least massive halo the two effects cross at $\sim5$\,yr, and by $10$\,yr the deeper sample is the more biased ($\sim57\%$ versus $\sim50\%$).
The practical conclusion is therefore not that depth has universally positive/negative impact, but that its value is set by whether sample size or velocity precision is the limiting factor. 
For the lowest-dispersion systems, where binary contamination matters most, spending exposure time on precision rather than on faint targets is the better trade at long baselines.

\subsection{Dependence on sample size and statistical regime}
\label{sec:disc_stat}

The impact of binary contamination is further modulated by the available sample size. 
We examine the individual realization of mock samples to examine the stochasticity explicitly, in addition to the median trend for different observing conditions.

\begin{figure}
    \centering
    \includegraphics[width=0.95\linewidth]{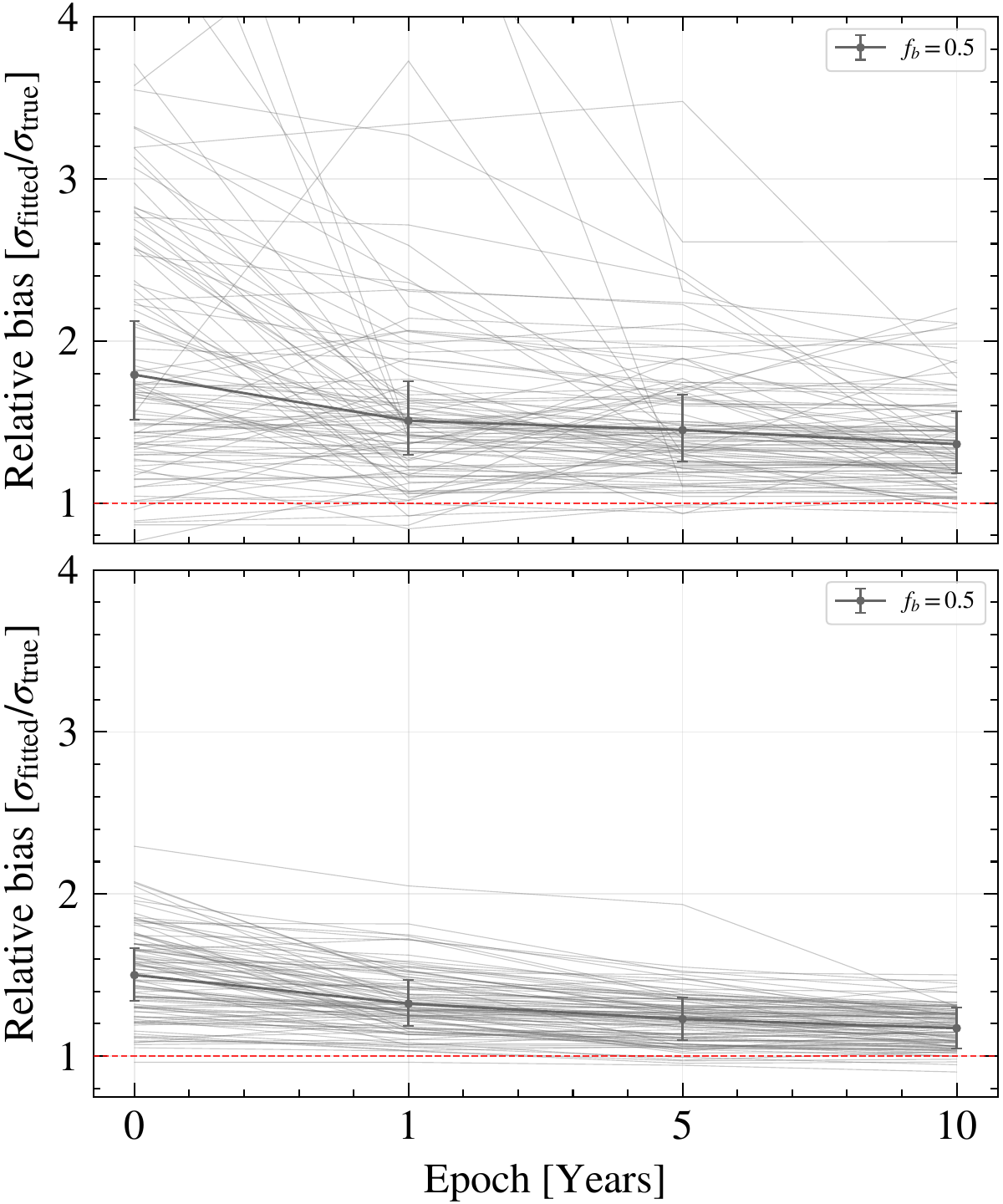}
    \caption{
    Comparison of the relative biases median trend and individual iterations for the $\sigma_{\rm true}=1.5$\,\kmsec\ halo with poor statistics (top; $50\%$ of $\sim30$ stars observed down to $G<20$ at each epoch) and good statistics (bottom; $80\%$ of $\sim100$ stars observed down to $G<21$).
    }
    \label{fig:stats_comp}
\end{figure}

For samples with more complete coverage ($80\%$ stars observed down to $G<21$) at each epoch, the posterior distributions for the velocity dispersion are well-resolved, and the effects of successive binary-rejection manifest as a clear and monotonic decrease in the inferred dispersion with increasing epoch baseline.
The robustness of the velocity dispersion estimates worsens as incomplete coverage ($50\%$ stars observed down to $G<20$) reduces the sample size overall.
When the effective sample size is reduced to $\lesssim20$ stars, comparable to the poorest-studied UFDs, the statistical uncertainties grow, and the recovered dispersion becomes increasingly sensitive to stochastic sampling effects. 
In this low-statistics regime, different random iterations from the same mock sample and analysis procedure can yield relative biases that differ by $\gtrsim100\%$ and do not necessarily decrease with longer multi-epoch baseline as new members are added. 
These fluctuations can obscure or even reverse the expected trend of decreasing dispersion with increasing baseline, as shown in Figure~\ref{fig:stats_comp}.
Importantly, this stochastic effect is \emph{not} captured by the quoted uncertainties on the dispersion fit.
Comparing the scatter across realizations with the median per-fit $1\sigma$ uncertainty for the two cases in Figure~\ref{fig:stats_comp}, the scatter exceeds the fitting uncertainty by a factor of $\sim6$ in the poor-statistics case and $\sim1.5$--$2.3$ in the good-statistics case.
Even when using the outlier-insensitive normalized median absolute deviation to quantify the scatter, so that the comparison is not driven by the few catastrophic realizations, the excess is still a factor of $\sim2$--$3$ and $\sim1.4$--$2.4$, respectively.
The reason is that the per-fit uncertainty describes only the sampling noise of the dispersion estimator given one particular set of measured velocities.
It carries no information about which stars happened to be drawn as binaries, or which stars happened to be visited at each epoch.
An error bar derived from a single campaign therefore understates the true uncertainty on the intrinsic dispersion of an UFD, and does so most severely where the sample is smallest.
This introduces an additional, and easily overlooked, systematic that complicates interpretation.
Together, these effects imply that, below a critical combination of galaxy mass and sample size, it becomes fundamentally difficult to draw concrete conclusions about the intrinsic velocity dispersion of any single system, regardless of observational baseline.

Furthermore, the stochasticity in the inferred dispersion also introduces challenges when one attempts to ``correct'' for the unidentified binaries, either through a binary-aware likelihood function or empirical corrections introduced later in this study. 
The naive assumption that the measured velocity dispersion should always be greater than the intrinsic dispersion may not hold true in the limit of small sample size and/or absence of multi-epoch observations for a majority of the star sample.
It is thus also crucial to carefully calibrate any correction to the measured velocity dispersion in those cases, as discussed in Section~\ref{sec:empirical_corr}.

\section{Link to observation: A case study with \bootes\,I}
\label{sec:link_obs}

The factors examined in previous sections can be combined to construct realistic forward models of both existing and future observing programs.
In this section, we compare the prediction to existing observations and literature velocity dispersion estimates of the UFD \bootes\,I over its $\sim17$-year baseline, driving the forward model with the real per-star observing record.
We then propose and study how an alternative observing strategy for \bootes\,I could have performed, providing recommendations for future observing campaign of UFDs.
Finally, we formulate an empirical correction that maps observed dispersions of \bootes\,I to its plausible intrinsic dispersion.
Together, these applications demonstrate that the forward model is not only useful for interpreting known systems, but also for translating heterogeneous observing histories into concrete expectations for dispersion recovery.

\subsection{Reproduction of literature results}
\label{sec:case_bootes1}

\bootes\,I is a well-studied ultra-faint Milky Way satellite with an exceptionally rich multi-epoch radial-velocity (RV) record spanning $\sim$17\,yr \citep{sandford26}. 
This makes it a natural validation case for our forward-modeling and cleaning pipeline. 
The intrinsic dispersion is large enough to be well-measured, while the heterogeneous cadence and instrumental setup provide a realistic test of the observational effects modeled here. 
Motivated by the recent Southern Stellar Stream Spectroscopic Survey ($S^{5}$) chemodynamical re-analysis that combines new AAT/AAOmega observations with archival spectroscopy to assemble a large member sample \citep{sandford26}, we forward model a ``\bootes\,I-like'' mock dataset, with sample size, per-epoch baseline, and uncertainties matching those from real observation.

Specifically, using the alternative mode described in Section~\ref{sec:methods}, we supply the member catalog and the per-epoch observing condition compiled by \citet{sandford26}.
Since compilation combines the $S^{5}$ AAT/AAOmega campaign with archival AAT/AAOmega, MMT/Hectochelle \citep{walker23} and VLT/GIRAFFE \citep{koposov11,jenkins21} spectroscopy, the per-epoch information is necessary to accurately reproduce the epochs at which each star was observed in the real campaign, together with that epoch's measured velocity uncertainty.
A total of over twenty nights of observation are included in the original observing schedule.
For a star with multiple measurements over consecutive nights within an observing run, the spectra are co-added and analyzed as a single measurement for the given run, with the systematic uncertainty included in the reported error \citep{sandford26}.
Additionally, we collapse the $2009$ VLT/GIRAFFE campaign, which accounts for over half of all individual measurements and occupies a $40$-day window, into a single combined epoch at $\Delta t = 1029$\,days following \citet{sandford26}.
For the MMT/Hectochelle campaign, co-adding the exposure does not change the total number of exposures for most stars because they are observed only once during a given observing run.
For the AAT/AAOmega campaign, most of the stars have $\sim3$ exposures across a multi-night observing run, so the co-add reduces the number of exposures for these stars.
Following the co-adding methodology, the mock observation and analysis are carried out over observing runs with at least a $15$-day gap in-between, resolving eight distinct epochs at $\Delta t = 0, 17, 348, 1034, 3251, 3658, 5151$ and $6158$\,days from the first MMT/Hectochelle measurements on 2006-05-08, spanning $\sim17$\,yr.
Because the compilation has already been placed on a common velocity scale, no inter-instrument zero-point offsets are injected.
In summary, the mock sample contains $127$ members with $301$ measurements in total, a mean of $2.4$ epochs per star, but the distribution is strongly skewed: only $\sim60\%$ of the stars are observed more than once and testable for velocity variability with a period longer than a few days.

For the true line-of-sight velocity dispersion of the mock sample, we adopt $\sigma_{\rm true}=4.0$\,\kmsec, as reported in the $S^{5}$ analysis \citep{sandford26}. 
We explore a grid of intrinsic binary fractions $f_{\rm b} = 0.1, 0.3, 0.5, 0.7, 0.9$.
We generate $200$ mock samples at each assumed $f_{\rm b}$, assuming identical per-epoch observing schedule and measurement uncertainties as described above.

We match the analysis to \citet{sandford26} as closely as the mock permits.
Velocity variables are identified with the same constant-velocity $\chi^{2}$ test, rejecting stars at $p<0.1$, and the surviving sample is fit as in Section~\ref{sec:mock_analysis}.
One element of the published procedure is not reproduced: \citet{sandford26} iterate the $p$-value cut jointly with the inter-instrument velocity zero-points, re-deriving the offsets after each rejection pass until both converge.
Because the compilation we used for compiling the mock samples had its zero-points corrected, we inject no instrument offsets into the mock. 
We thus apply the binary cut once without iteration for potential instrument zero-points offset.
We retain the iterative $3\sigma$ velocity clip as a proxy for the membership selection, noting that the input sample is already members-only, so the clip removes residual velocity outliers rather than performing the membership cut itself.

\begin{figure*}
    \centering
    \includegraphics[width=0.49\linewidth]{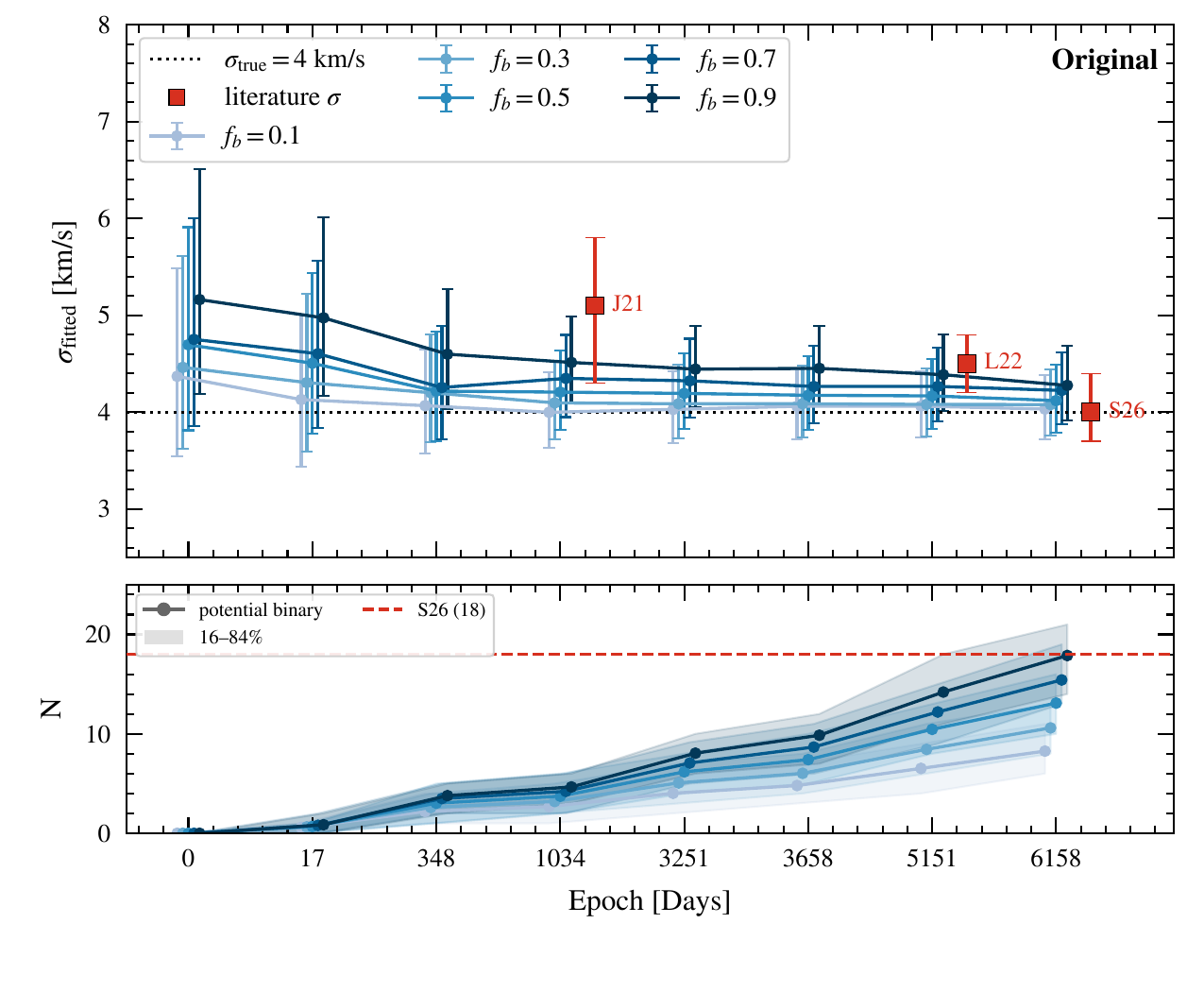}
    \includegraphics[width=0.49\linewidth]{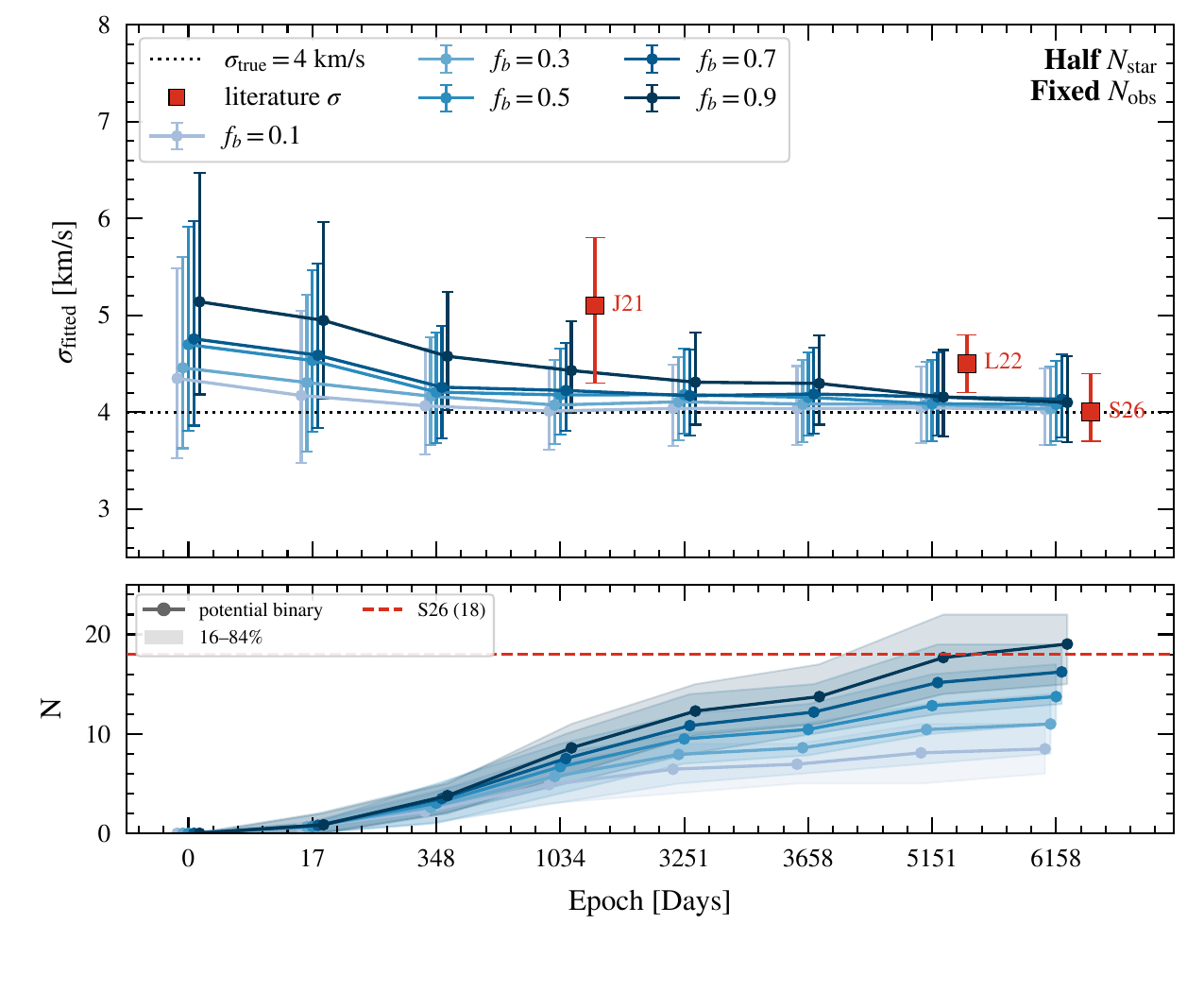}
    \caption{
    The recovered velocity dispersion for \bootes\,I as a function of observing baseline (top) and the number of stars flagged as potential binaries (bottom), forward-modeled through the per-epoch observing schedule of the \citet{sandford26} compilation.
    The left panels show the observing record as executed in the literature.
    The right panels show the same $301$ measurements redistributed onto half of the sample (Section~\ref{sec:strategy_pred}), preserving the number of observations at every epoch, so that every star becomes testable for velocity variability (see Section~\ref{sec:strategy_pred}).
    Colors denote the assumed intrinsic binary fraction in both panels.
    In the top panels, the error bars are the median fitted dispersion uncertainties over $200$ realizations.
    The dotted line is the assumed intrinsic dispersion ($4.0$\,\kmsec) from \citet{sandford26}.
    Red squares mark published dispersions from \citet{jenkins21} (J21), \citet{longeard22} (L22) and \citet{sandford26} (S26).
    Only \citet{sandford26} analyzes the same compilation reproduced here.
    In the bottom panels, the shaded regions are the $16$-$84$th percentiles of the number of potential binaries identified over $200$ realizations.
    The dashed line indicates the number of the potential binaries flagged in the real \citet{sandford26} sample.
    }
    \label{fig:bootesi_cleaning}
\end{figure*}

The resulting measured velocity dispersion tracks are shown in the left panel of Figure~\ref{fig:bootesi_cleaning}.
The behavior seen in the literature is reproduced: single-epoch estimates are inflated well above the input dispersion, and the inferred value falls as the baseline lengthens and velocity variables are progressively identified.
At $f_{\rm b}=0.5$, for example, the cleaned dispersion declines from $\sim4.7$\,\kmsec\ at the first epoch to $\sim4.1$\,\kmsec\ at $6158$\,days, against an input $\sigma_{\rm true}=4.0$\,\kmsec, while the number of stars flagged as velocity variables grows to $\sim13$.
Most of the decline is complete by $\Delta t\sim1000$\,days, once the VLT campaign is included. 
The remaining $\sim5000$\,days of baseline removes comparatively little.
The residual excess at the final epoch is $\sim+0.12$\,\kmsec\ at $f_{\rm b}=0.5$ and $\sim+0.28$\,\kmsec\ at $f_{\rm b}=0.9$, so even a $17$-yr record with this cadence does not fully remove the binary contribution.

Comparison with the published dispersions should be interpreted with caution.
Our schedule reproduces the full compilation, so the \citet{sandford26} value of $4.0^{+0.4}_{-0.3}$\,\kmsec\ at $6158$\,days is the most direct comparison and consistent with our mock analysis for all $f_{\rm b}$ cases.
Additionally, \citet{sandford26} identified 18 binary candidates, which is consistent with the $f_{\rm b}\gtrsim0.7$ cases, as shown in the bottom left panel of Figure~\ref{fig:bootesi_cleaning}.
This shows the mock can also be used to provide qualitative constraints on the binary fraction of the system. 
The other two studies use different data and different pipelines.
\citet{jenkins21} measured $5.1^{+0.7}_{-0.8}$\,\kmsec\ from VLT/GIRAFFE alone, a single $40$-day campaign in $2009$ that we deliberately collapse into one epoch, so their measurement samples a far shorter effective baseline than its position on the axis suggests.
\citet{longeard22} obtained $4.5^{+0.3}_{-0.3}$\,\kmsec\ from an unspecified selected subset of the AAT/AAOmega and VLT/GIRAFFE, and the dynamical modeling allows a non-zero velocity gradient.
We therefore plot all three for orientation but treat the agreement with these two studies as qualitative only.

Overall, driving the forward model with the real observing record reproduces the literature trend without any tuning of the coverage.
In Appendix~\ref{sec:app_ufds}, we apply the same procedure to four further UFDs with published per-star observing records.
Those systems' observation baselines are much shorter than \bootes\,I's, and we discuss whether a single additional epoch can provide a measurable improvement in removing the binaries.

\subsection{Alternative observing strategy}
\label{sec:strategy_pred}

Having reproduced the observing record, we next examine what the \emph{same} observing effort would have delivered if allocated differently, applying the results from Section~\ref{sec:sparse_visit}.
We therefore construct an alternative schedule that holds the total budget fixed.
All $301$ measurements are retained, and the number of stars observed at every individual epoch is preserved, so the alternative campaign is approximately identical in telescope time, epoch structure, and per-epoch multiplexing.
The only change is which stars are targeted: the observations are redistributed onto the $69$ best-covered stars, half the original sample.
Each moved observation keeps the uncertainty of the epoch it belongs to, so precision is not improved beyond what was available.
For consistency, the original and alternative schedules are evaluated on the same set of 200 realizations for a given binary fraction case.

We tested two ways of performing the redistribution: (1) concentrating the freed observations on the already best-covered stars, and (2) raising the floor so that every retained star reaches at least two epochs.
The two differ only marginally. 
The paired per-realization difference is $0.02$--$0.06$\,\kmsec, with option (1) performing slightly worse of the two at most binary fractions, but with both an order of magnitude below the gain from the redistribution itself. 
We therefore show the second case, which is both marginally better and more flexible to execute in real observations. 
The redistribution raises the mean from $2.4$ to $4.4$ epochs per star and, more importantly, eliminates the single-epoch population entirely. 
All but one star in the alternative sample has between three and seven measurements, and all $69$ are testable for long-period velocity variability.
The right panel of Figure~\ref{fig:bootesi_cleaning} shows the consequence.
Up to $f_{\rm b}=0.9$, the final-epoch dispersion falls to $\lesssim4.10$\,\kmsec\ at the last epoch.
The alternative campaign also converges earlier, tracking the input dispersion from $\Delta t\sim3000$\,days onward rather than plateauing $\sim0.25$\,\kmsec\ above it for $f_{\rm b}$ up to 0.7.
Even in the most extreme $f_{\rm b}=0.9$ case, the fitted dispersion converges at $\Delta t\sim5000$\,days.
The cost is statistical, halving the sample widens the typical fitted dispersion uncertainties by roughly $\sqrt{2}$, visible as the larger error bars in the right panel of Figure~\ref{fig:bootesi_cleaning}.
Thus, the alternative observing strategy trades statistical uncertainty for systematic uncertainty from binaries. 


For accurate bias-free velocity dispersion, it is thus clear that, for \bootes\,I, with the observing time already spent, targeting half as many stars twice as often would have removed roughly half of the residual binary bias.
The broader lesson matches Section~\ref{sec:sparse_visit}.
Single-epoch measurements contribute to the dispersion fit but carry no information for binary rejection, so a campaign that accumulates many singly-observed members buys sample size at the expense of accuracy.
Where the intrinsic dispersion is the quantity of interest, repeat coverage of a smaller member sample is the better investment.

\subsection{Empirical correction relations}
\label{sec:empirical_corr}

We next explore the feasibility of constructing empirical correction relations from observed dispersion to intrinsic dispersion for UFDs.
As shown in previous sections, reaching the intrinsic dispersion of a system like \bootes\,I requires not merely a long baseline but a sample in which most members are observed more than once.
With the existing observing data, even a $17$-yr baseline is expected to leave a residual bias in the fitted velocity dispersion for \bootes\,I.
For systems without such baselines or observing resources, empirical corrections provide a practical way to quantify the remaining binary-induced uncertainty in the observed dispersion \citep[e.g.,][]{arroyopolonio26}.

Constructing such an empirical correction with our forward model follows a procedure very similar to what was done so far, with one important difference.
For existing observations, our tests with \bootes\,I (as well as the four UFDs of Appendix~\ref{sec:app_ufds}) to project the derived velocity dispersions are achieved by assuming some fixed intrinsic dispersion and varying observing strategies.
For the empirical correction, we instead ask the inverse question: given a fixed observing strategy and a measured dispersion, what range of intrinsic dispersions could plausibly have produced that measurement?
In other words, the correction is conditioned on both the observing conditions and the observed velocity dispersion.
With our forward modeling tool, this means generating an ensemble of mock samples and associated analyses with a fixed observing strategy and varying intrinsic velocity dispersion.
In particular, for any given observing conditions/strategies employed, the correction must factor in not only the median trend but also the expected amount of variance in individual realizations for the given observing condition, as shown in Section~\ref{sec:disc_stat}. 
For UFDs with limited statistics ($\lesssim30$ stars), the large variance in derived dispersion from a fixed intrinsic dispersion means that there would equally exist a significantly large range of intrinsic dispersion that can map to a fixed derived dispersion.

\begin{figure}
    \centering
    \includegraphics[width=\linewidth]{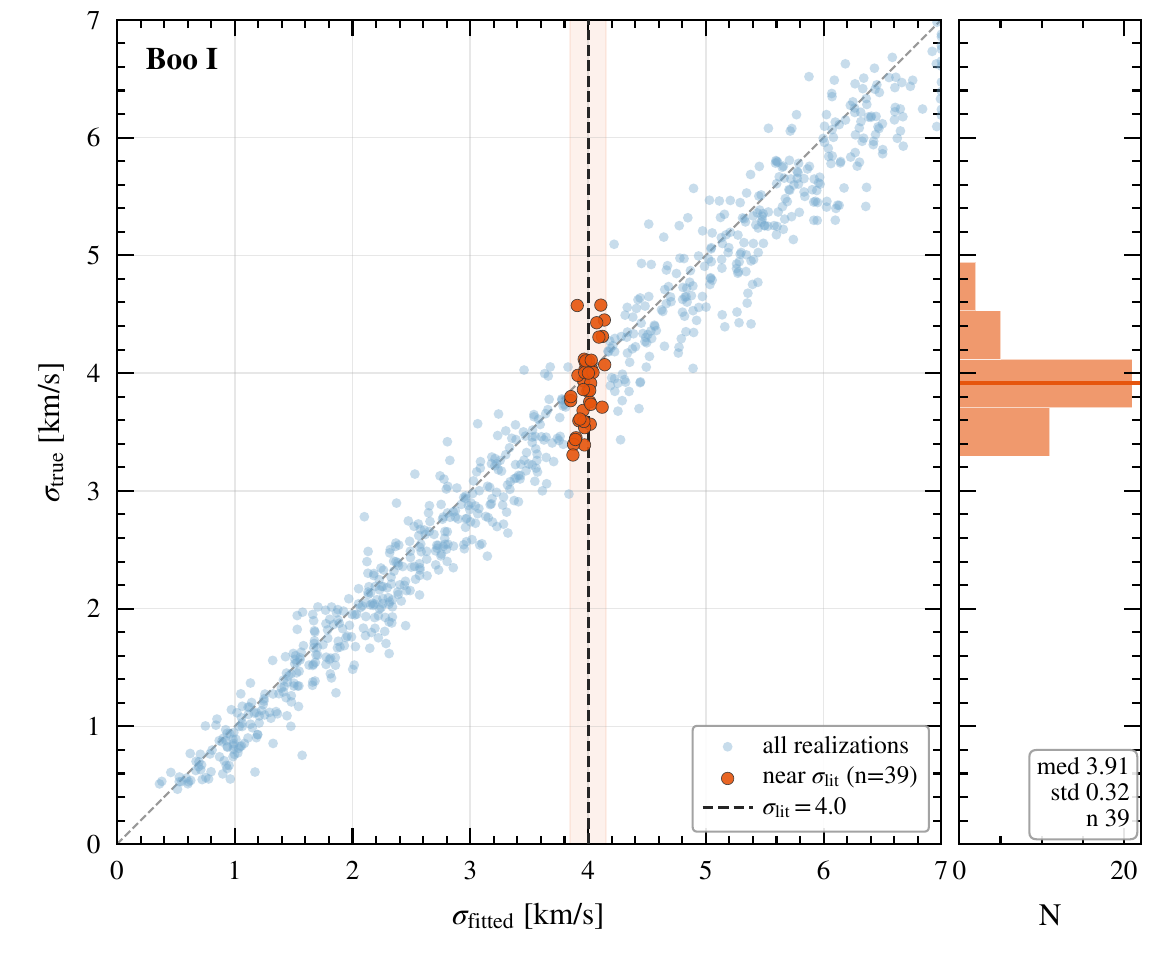}
    \caption{
    The empirical correction for \bootes\,I at the final observing epoch, from an ensemble of realizations forward-modeled through the real \citet{sandford26} schedule and an assumed $f_{\rm b}=0.5$.
    Each blue point is one realization; the horizontal axis is the fitted dispersion and the vertical axis the intrinsic dispersion that produced it.
    The gray dashed line marks the unbiased case.
    Orange points are the $39$ realizations whose recovered dispersion falls within $0.15$\,\kmsec\ of the published value \citep{sandford26}, and the side histogram shows the distribution of their intrinsic dispersions.
    The median of that distribution is the corrected dispersion and its width the binary-induced uncertainty floor for this observing strategy.
    }
    \label{fig:empri_corr}
\end{figure}

We therefore generate an ensemble of $800$ \bootes\,I-like mock samples, following Section~\ref{sec:case_bootes1}, but with intrinsic velocity dispersion sampled uniformly in linear space and a fixed binary fraction of $0.5$.
The uniform sampling in velocity dispersion is arbitrary and does not map to physically motivated sampling in properties such as the galaxy halo mass, the resulting distribution of plausible intrinsic dispersions should be interpreted as qualitative and conditional on the prior adopted.
Nonetheless, for the purpose of mapping the possible range of intrinsic dispersion, we expect the mapping to be robust as long as the ensemble is dense enough to fully sample the variance around the median trend.
Figure~\ref{fig:empri_corr} shows the ensemble in intrinsic versus measured dispersion space at the final epoch, together with the subset of realizations whose recovered dispersion lands within some small window of the published value. 
Ideally the window should be a $\delta$-function centered exactly on the published value, given the goal of the study.
In practice, we adopt an arbitrary window size of $0.15$\,\kmsec, such that it is narrow enough to not be driven by the increasing trend of the iteration ensemble distribution, but wide enough to include tens of mock iterations to sample the $\sigma_{\rm true}$ distribution.
The results do not qualitatively depend on the choice of the window size.
We examine the mapping and compare to published values for only the last epoch, where our analysis is matched to that of \citet{sandford26}.

Of the ensemble, $39$ realizations reproduce the published dispersion at $f_{\rm b}=0.5$, and their intrinsic dispersions center on $\sigma_{\rm true}=3.91$\,\kmsec\ with a standard deviation of $0.32$\,\kmsec.
The measured value therefore overestimates the intrinsic dispersion by $\sim2\%$ for this campaign, if the true $f_{\rm b}=0.5$.
The width of the mapped distribution ($0.32$\,\kmsec) can be interpreted as a floor on the binary-induced uncertainty, conditioned on this observing strategy, which should be carried alongside the formal statistical error from the standard velocity dispersion fitting procedure.
The result is only weakly sensitive to the assumed binary fraction, giving $3.94\pm0.24$\,\kmsec\ at $f_{\rm b}=0.3$ and $3.79\pm0.37$\,\kmsec\ at $f_{\rm b}=0.7$. 
The inferred correction is stable while its uncertainty grows with $f_{\rm b}$, as expected when more binaries remain unidentified.

Thus, to first order, the empirical correction shown in Figure~\ref{fig:empri_corr} provides a straightforward calibration from measured to plausible intrinsic dispersion for a specified observing strategy.
This forward-modeling approach does not replace a full posterior inference of the intrinsic dispersion, binary fractions, or binary population parameters (e.g., \citealt{spencer17,spencer18,minor19,arroyopolonio23}). 
Instead, it offers a complementary and computationally efficient method for propagating binary-induced systematics into existing dispersion measurements, especially when the observing strategy is too heterogeneous for a simple analytic likelihood.
In this sense, the empirical correction relation is most useful as a fast, observation-conditioned calibration layer that can guide both interpretation of existing UFD data and the design of future monitoring campaigns.

\section{Future steps} 
\label{sec:limit}

The mock samples and analyses presented in this work are designed as a flexible first implementation of a forward-modeling framework for quantifying binary-induced biases in dwarf galaxy dynamical mass estimates. 
Rather than attempting to model every astrophysical and observational effect simultaneously, we have focused on isolating the dependence of the inferred velocity dispersion on binary fraction, galaxy mass scale, observational baseline, measurement precision, and epoch coverage. 
This controlled setup makes the main trends physically interpretable, while also identifying several natural extensions that can make the framework increasingly realistic for direct application to individual observed systems.

A first extension concerns the treatment of internal binary evolution. 
The current implementation follows the kinematic imprint of binaries under prescribed orbital-parameter distributions \citep{dupuennoy91,moe17}. 
These distributions are constructed primarily based on solar-metallicity ZAMS stars, whereas most stars observed in dwarf galaxies are metal-poor giants.
The low metallicity ([Fe/H]$\lesssim-2$) of the stellar population is correlated with higher close binary fraction \citep[e.g.,][]{moe19}, whereas interacting binary phases on the giant branch, such as Roche-lobe overflow, modify stellar masses, luminosities, survivability, and orbital properties \citep[e.g.,][]{breivik20}. 
Incorporating parametrized binary-evolution prescriptions would therefore allow the same framework to connect the radial-velocity variability explored in this work with photometric signatures of interacting or evolved binaries.
A related caveat applies to the mock samples used here.
Because stellar masses are drawn from the initial mass function and binaries are paired afterwards, the pairing step is not restricted to the masses that survive to the age of an ancient stellar population; a small tail of companions is therefore more massive than any star that should remain on the main sequence in a $13.5$\,Gyr system.
These companions produce reflex velocities larger than a real UFD would exhibit, and the data-driven clip described in Section~\ref{sec:mock_analysis} removes the great majority of them.
The residual biases we report should nonetheless be read as conservative upper bounds.
Eliminating this tail self-consistently requires sampling a system mass function and synthesizing coeval binary systems directly, rather than pairing independently sampled single stars, which we defer to future work.

A second extension is to broaden the range of galaxy structural models. 
For the controlled experiments in this study, the mock systems are initialized with simple dark-matter and stellar profiles and an approximately fixed global mass-to-light ratio, allowing us to isolate how binary contamination scales with the intrinsic velocity dispersion. 
Future applications can relax this setup by adopting system-specific mass-to-light ratios, multiple stellar populations, or clustered stellar components. 
Because the simulator is already modular in its treatment of stellar and dark-matter profiles, these additions can be incorporated without changing the core multi-epoch observing framework.

A third extension is the inclusion of foreground contamination. 
The present experiments intentionally focus on member stars in order to isolate the effect of binary orbital motion on the recovered dispersion. 
For direct comparisons to individual Milky Way satellites, foreground stars can be added as a separate, location-dependent component, since the contaminant population depends on sky position, distance, proper-motion selection, photometric cuts, and metallicity information. 
Adding such a component would enable the framework to test the interaction between binary rejection, membership selection, and velocity-dispersion inference in realistic survey fields.

The forward-modeling approach also opens a clear path toward simulation-based inference, although additional computational development is needed. 
The current mock-generation procedure relies on extensive stochastic sampling and repeated mock analyses, which is well suited for exploring controlled observing strategies and constructing empirical correction relations, but is not yet optimized for full posterior inference from observed data. 
Future development will focus on accelerating the simulator, building reusable mock libraries or emulators, and enabling likelihood-free, likelihood-based, or hierarchical inference directly conditioned on observed multi-epoch datasets.

Finally, the analysis module can be made more closely matched to modern observational pipelines. 
In this work, we intentionally adopt a simple and transparent binary-cleaning procedure, which makes the origin of the measured biases easy to interpret across the mock grid. 
Future versions can incorporate proper motions and photometry for probabilistic membership selection, jointly fit binary fraction and intrinsic velocity dispersion using the full multi-epoch likelihood, and interface with more sophisticated Jeans modeling or machine-learning approaches when the available sample size is large enough. 
These additions would allow the same forward-modeling framework to test not only the observing strategy itself, but also the robustness of different analysis choices applied to the same underlying galaxy.

Taken together, these extensions would move the framework from controlled experiments toward a fully observation-conditioned inference tool. 
The key advantage of the present approach is that each added layer can be tested independently before being combined into a complete end-to-end model for individual dwarf galaxies.

\section{Conclusion}
\label{sec:conclusion}

In this study, we introduce the BOS, a flexible forward-modeling framework for quantifying how unresolved binary stars bias velocity-dispersion-based dynamical mass estimates of UFDs.
\footnote{The code is available as a repository on GitHub (\url{https://github.com/xou-mit/binary-observation-simulator}).}
We construct mock dwarf-galaxy samples spanning three halo masses, $M_{\rm halo}\sim10^{4.5},10^{5.5},10^{6.5}\,\msun$, corresponding to intrinsic dispersions of $\sigma_{\rm true}\sim0.75,1.5,3.5$\,\kmsec, and binary fractions from $0.1$ to $0.9$.
By applying mock multi-epoch binary rejection and velocity-dispersion inference, we isolate how the inferred dispersion bias depends on galaxy mass scale, binary population, velocity precision, sample size, repeat-observation completeness, and observing baseline.

Our main conclusions are as follows:

\begin{itemize}

\item Binary contamination is most severe for the dynamically coldest systems (Section~\ref{sec:dep_galprop}). 
For the lowest-mass halo in our grid, with $\sigma_{\rm true}\sim0.75$\,\kmsec, single-epoch or short-baseline measurements overestimate the dispersion by $\sim20$--$320\%$, corresponding to an absolute bias up to $\sim2$\,\kmsec. 
Multi-epoch monitoring reduces the bias, but the required baseline depends strongly on the intrinsic dispersion. 
Even a 10-yr baseline can leave $\sim10$--$120\%$ relative bias when repeat coverage is incomplete for the $\sigma_{\rm true}\sim0.75$\,\kmsec\ case.

\item For observing strategy, repeat coverage is the single most reliable lever, while the value of observational depth depends on the system (Section~\ref{sec:sparse_visit}).
In the $f_{\rm b}=0.5$ comparison, increasing the repeat-observation fraction from $50\%$ to $80\%$ reduces the 10-yr relative bias in the velocity dispersion of the lowest-mass halo from $\sim50\%$ to $\sim35\%$. 
Pushing from $G<20$ to $G<21$ instead trades velocity precision for sample size, raising the typical uncertainty from $\sim0.5$ to $\sim2$\,\kmsec. 
The larger sample wins for the two more massive halos, but for the coldest system the added uncertainty leaves more binaries unidentified and the deeper sample ends up the more biased of the two by the end of the $10$\,yr observation baseline.

\item Small-number statistics remain a limiting factor for the faintest systems (Section~\ref{sec:disc_stat}). 
When the effective sample size falls to $\lesssim20$ stars, individual mock realizations can differ in relative bias by $\gtrsim100\%$, and the inferred dispersion does not always decrease monotonically with increasing baseline. 
In general, these stochastic effects are not captured by the statistical uncertainties on the dispersion fit.
For such systems, empirical corrections or binary-aware likelihoods should propagate the realization-to-realization scatter, rather than assuming that the observed dispersion is always an upper limit.

\item Driving the forward model with a real observing record reproduces the literature behavior without tuning and highlights repeat coverage is what limits our ability to constrain binary-induced bias.
Using the per-star schedule of the \citet{sandford26} \bootes\,I compilation, we qualitatively reproduced not only the literature dispersion measurements but also the number of potential binaries identified (Section~\ref{sec:case_bootes1}).
This highlights that the tool can be used both for empirically calibrating the fitted dispersion and constraining the binary fraction.
Redistributing the measurements at no additional observing resource cost, we show that the measured dispersion converges faster to the true value regardless of binary fraction (Section~\ref{sec:strategy_pred}).

\item The binary fraction is not the dominant uncertainty once it is moderately high. 
Increasing $f_{\rm b}$ has a diminishing effect once $f_{\rm b}\gtrsim0.5$, especially for systems with $\sigma_{\rm true}\gtrsim1.5$\,\kmsec. 
This saturation suggests that empirical correction relations conditioned on the observing strategy are feasible even when the true binary fraction is not precisely known.
For the \bootes\,I empirical correction with $f_{\rm b}=0.5$, the plausible intrinsic dispersion inferred at the final epoch centers on $\sigma_{\rm true}\simeq3.91$\,\kmsec\ with a scatter of $\sim0.32$\,\kmsec, and shifts by less than $0.2$\,\kmsec\ when $f_{\rm b}$ is varied between $0.3$ and $0.7$ (Section~\ref{sec:empirical_corr}).

\end{itemize}

Taken together, these results show that unresolved binaries remain a central systematic for dynamical mass estimates of the faintest and coldest dwarf galaxies.
The bias is not controlled by a single physical parameter, but by the combination of intrinsic dispersion, binary population, sample size, radial-velocity precision, epoch baseline, and repeat-observation completeness.
Because the inferred dynamical mass scales directly with the square of the velocity dispersion, even modest residual inflation can translate into substantial mass overestimates, affecting interpretations of the low-mass end of the galaxy-halo connection and tests of dark matter models.

The practical takeaway is therefore that future UFD spectroscopy should prioritize precise, repeated measurements of a well-defined bright member sample, especially for the lowest mass systems with expected $\sigma_{\rm true}\lesssim1$\,\kmsec.
For systems lacking sufficiently complete long-baseline monitoring, the safer path is to forward model the actual observing cadence and construct empirical or hierarchical corrections conditioned on the available data, rather than applying a universal binary correction or relying only on hard binary cuts.
At a fixed assumed binary fraction of $0.5$, the inferred overestimate of the published dispersion ranges from $\sim2\%$ for \bootes\,I, whose $127$ members are sampled over eight epochs, to $\sim24\%$ for Car\,II, whose $18$ members are sampled over five epochs (Appendix~\ref{sec:app_ufds}). 
A factor of four spread at fixed binary fraction is not something a single universal correction can absorb, and it is recoverable only by modeling the cadence that produced each measurement.

Several extensions are needed before this framework can be applied as a full inference tool to the Milky Way satellite population.
Future versions will include more realistic target selection, foreground contamination, instrument-specific zero-point systematics, detailed binary evolution, and joint modeling of photometric and astrometric information.
The present implementation nevertheless provides a necessary first step: a controlled forward model that connects binary populations, observing strategy, and velocity-dispersion bias in a way that can be customized for individual galaxies.
Such forward modeling will be essential for turning heterogeneous multi-epoch spectroscopic datasets into robust dynamical mass constraints for the lowest-mass galaxies.

\section{Software and third party data repository citations} \label{sec:cite}

\software{%
matplotlib \citep{hunter07},
numpy \citep{vanderwalt11},
scipy \citep{jones01},
emcee \citep{foremanmackey13}, 
astropy \citep{astropy:2013,astropy:2018}, 
dyad \citep{gration25}, and
StellarTracks.jl \citep{garling25}.}

\begin{acknowledgments}

XO and NK acknowledge support from the NSF-Simons AI Institute for Cosmic Origins which is supported by the National Science Foundation under Cooperative Agreement 2421782 and the Simons Foundation award MPS-AI-00010515.
LN is supported by the Sloan Fellowship, the NSF CAREER award 2337864, NSF award 2307788, and by the NSF award PHY2019786 (The NSF AI Institute for Artificial Intelligence and Fundamental Interactions, \href{http://iaifi.org/}{http://iaifi.org/}). LN also gratefully acknowledges the continued support of the Adam J. Burgasser Endowed Chair of Astrophysics at MIT, which sustained many long hours of writing and revision of this manuscript.
The authors acknowledge Research Computing at The University of Virginia for providing computational resources and technical support that have contributed to the results reported within this publication. URL: \href{https://rc.virginia.edu}{https://rc.virginia.edu}

\end{acknowledgments}

\clearpage

\bibliography{xou,xou_add}{}
\bibliographystyle{aasjournalv7}

\renewcommand{\thefigure}{\thesection\arabic{figure}}
\renewcommand{\thetable}{\thesection\arabic{table}}

\appendix

\section{Sensitivity to the binary-cleaning scheme}
\label{sec:app_cleaning}
\setcounter{figure}{0}
\setcounter{table}{0}

The relative biases reported in this paper are a property of the observing strategy \emph{and} of the analysis applied to it. 
Because the second is a choice rather than a measurement, we quantify here how much it matters, by repeating the entire mock grid with the two cleaning schemes described in Section~\ref{sec:mock_analysis} applied to the \emph{same} underlying mock samples.

The two schemes are (i) a \emph{fixed window}: a single $5\sigma$ membership cut constructed from an assumed intrinsic dispersion, followed by a per-star $\chi^2$ consistency test on repeat visitors; and (ii) the \emph{guess-free clip} adopted in this work: the same $\chi^2$ test followed by an iterative $3\sigma$ velocity clip whose threshold scales with the fitted dispersion.
For scheme (i) we set the assumed dispersion equal to the intrinsic dispersion of each mock galaxy, which is the most favorable choice possible and is not information available to a real observer.

\begin{deluxetable*}{cccccccccc}
\tablecaption{Median relative bias $\sigma_{\rm fitted}/\sigma_{\rm true}$ under the two cleaning schemes at $50\%$ coverage, for two observational depths (see Section~\ref{sec:sparse_visit}).
``Fixed'' uses an assumed intrinsic dispersion set equal to the true value of each mock galaxy; ``clip'' is the guess-free scheme adopted in the main text.
\label{tab:cleaning}}
\tablehead{
\colhead{} & \colhead{} & \multicolumn{4}{c}{$G<20$} & \multicolumn{4}{c}{$G<21$} \\
\cline{3-6} \cline{7-10}
\colhead{} & \colhead{} & \multicolumn{2}{c}{Epoch 0} & \multicolumn{2}{c}{Epoch 3} &
\multicolumn{2}{c}{Epoch 0} & \multicolumn{2}{c}{Epoch 3} \\
\colhead{$\sigma_{\rm true}$ [\kmsec]} & \colhead{$f_{\rm b}$} &
\colhead{fixed} & \colhead{clip} & \colhead{fixed} & \colhead{clip} &
\colhead{fixed} & \colhead{clip} & \colhead{fixed} & \colhead{clip}
}
\startdata
$0.75$              & 0.5 & 2.19 & 2.76 & 1.89 & 1.50 & 2.58 & 2.21 & 2.18 & 1.57 \\
\hline
$1.5$             & 0.5 & 1.54 & 1.72 & 1.36 & 1.33 & 1.60 & 1.48 & 1.44 & 1.26 \\
\hline
$3.5$              & 0.5 & 1.29 & 1.42 & 1.20 & 1.14 & 1.34 & 1.20 & 1.19 & 1.07 \\
\enddata
\end{deluxetable*}
\onecolumngrid

Table~\ref{tab:cleaning} shows that the two schemes disagree in opposite directions at the two ends of the baseline. 
At the first epoch the fixed window is the more accurate of the two, and increasingly so as the binary fraction rises: for the least massive halo at $f_{\rm b}=0.5$ it returns $2.19$ against the clip's $2.76$. 
This is not a failure of the clip but a consequence of the information each scheme uses. 
With a single epoch the $\chi^2$ test is inert, so the clip must estimate its own rejection threshold from a sample whose dispersion is already inflated by the binaries it is trying to remove. 
The fixed window escapes this only because it has a good initial guess of the intrinsic dispersion.

The ordering reverses once a baseline exists. 
By the final epoch the clip is the more accurate scheme in every configuration.
For the least massive halo at $f_{\rm b}=0.5$ the fixed window method returns $1.89$ against the clip's $1.50$.
We further verify that the fixed window result is strongly sensitive to the assumed dispersion it is given. 
A narrower initial window can improve the relative bias but risk removing true member stars in the high-amplitude tail of the velocity dispersion.
A scheme whose answer depends this strongly on a quantity the observation is meant to measure is not a safe basis for a strategy recommendation.

The choice of cleaning does not merely rescale the biases; it can change the strategic conclusion drawn from them as shown in Table~\ref{tab:cleaning}. 
Under the fixed window at $f_{\rm b}=0.5$ and the final epoch, going from $G<20$ to $G<21$ raises the relative bias from $1.89$ to $2.18$ for the least massive halo and from $1.36$ to $1.44$ for the intermediate one, and leaves the most massive unchanged within the scatter ($1.20$ to $1.19$).
Read on its own, that is a clean recommendation to spend exposure time on velocity precision rather than on faint targets.
Under the guess-free clip the same comparison gives $1.50 \to 1.57$, $1.33 \to 1.26$ and $1.14 \to 1.07$: the deeper sample is now the better choice for the two more massive halos, and only the least massive retains the original ordering.
The two schemes thus support opposite advice on how to spend telescope time, from identical data.
A strategy recommendation therefore cannot be quoted independently of the analysis that will be applied to the data.
We encourage future forward-modeling studies to state the cleaning scheme explicitly and, where a strategy conclusion is drawn, to verify that it is stable against a reasonable alternative.
Additionally, real observational datasets need to deal with Milky Way contamination. 
For studies that forward-model such contamination, the membership selection criteria are also expected to alter any conclusion one may draw.
We plan to explore the effect quantitatively as we incorporate contamination models into the tool described in this work.

\section{Additional comparison with literature UFD campaigns}
\label{sec:app_ufds}
\setcounter{figure}{0}
\setcounter{table}{0}

Section~\ref{sec:link_obs} applies the framework to \bootes\,I with per-star observing condition from \citet{sandford26}. 
Here we extend the comparison to four further systems for which a published per-star observing record is available.
The four systems are Centaurus\,I and Eridanus\,IV \citep{heiger26}, Pictor\,II \citep{pace25}, and Carina\,II \citep{li18}. 
They span a useful range of campaign designs: $13$-$34$ members, four to five epochs, and total baselines from $113$ to $747$ days.

\begin{figure*}
    \centering
    \includegraphics[width=0.48\linewidth]{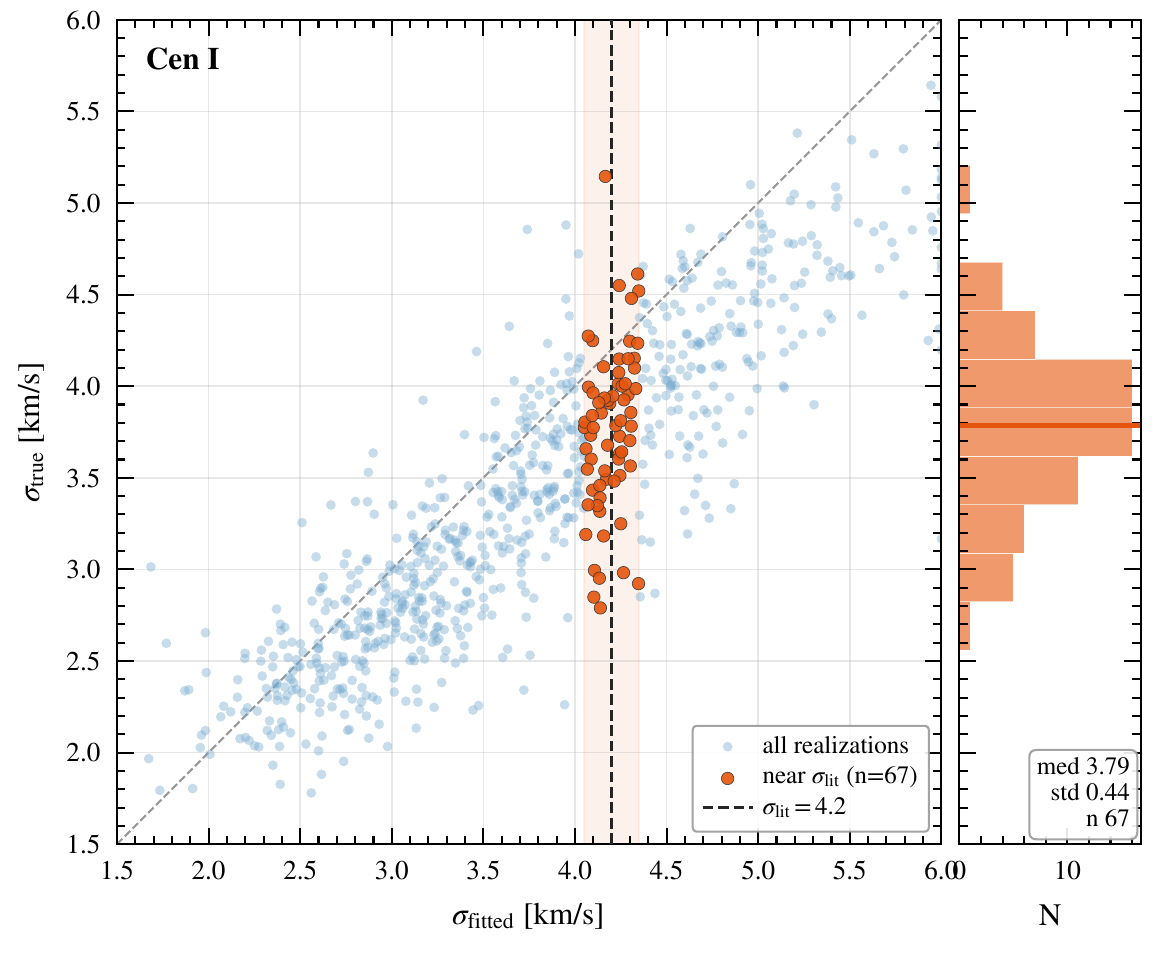}
    \includegraphics[width=0.48\linewidth]{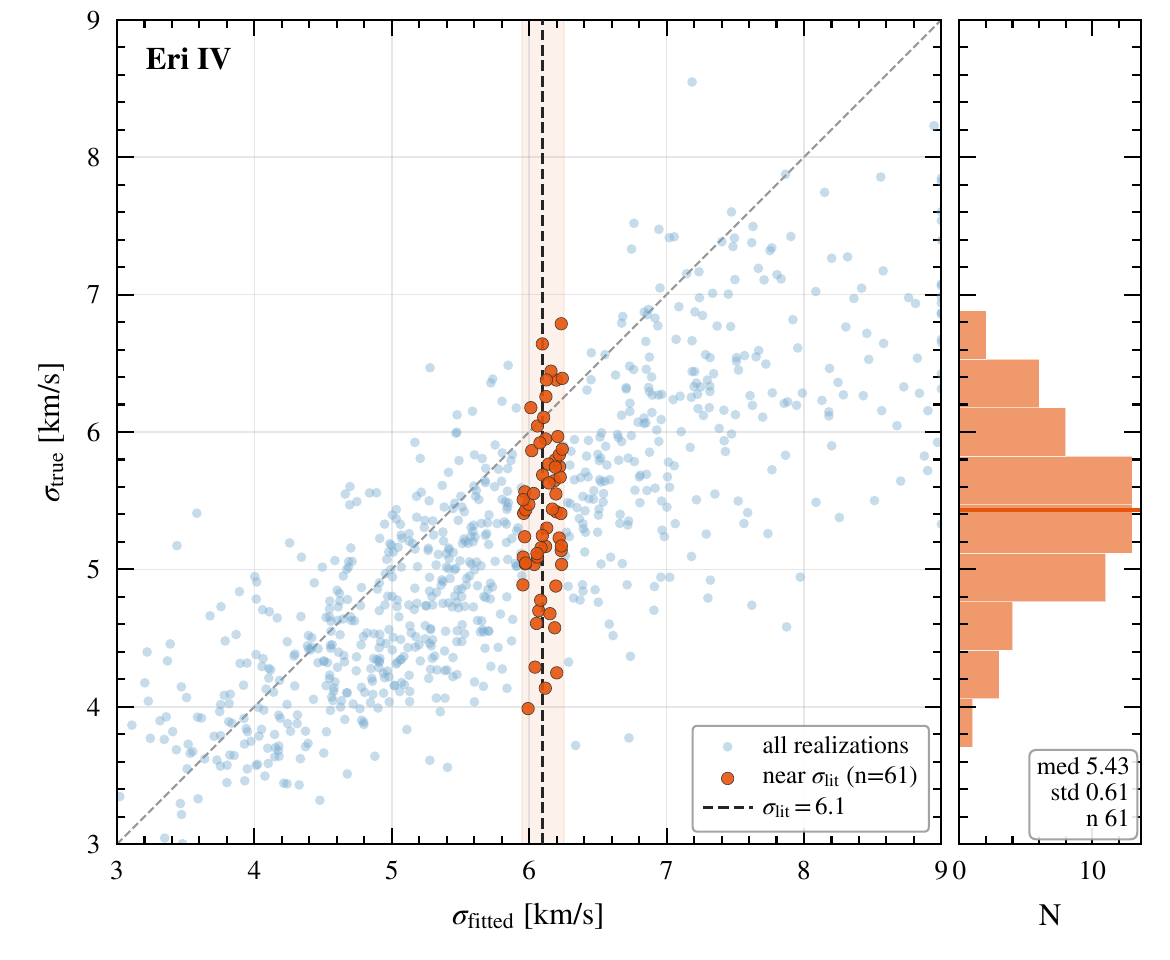} \\
    \includegraphics[width=0.48\linewidth]{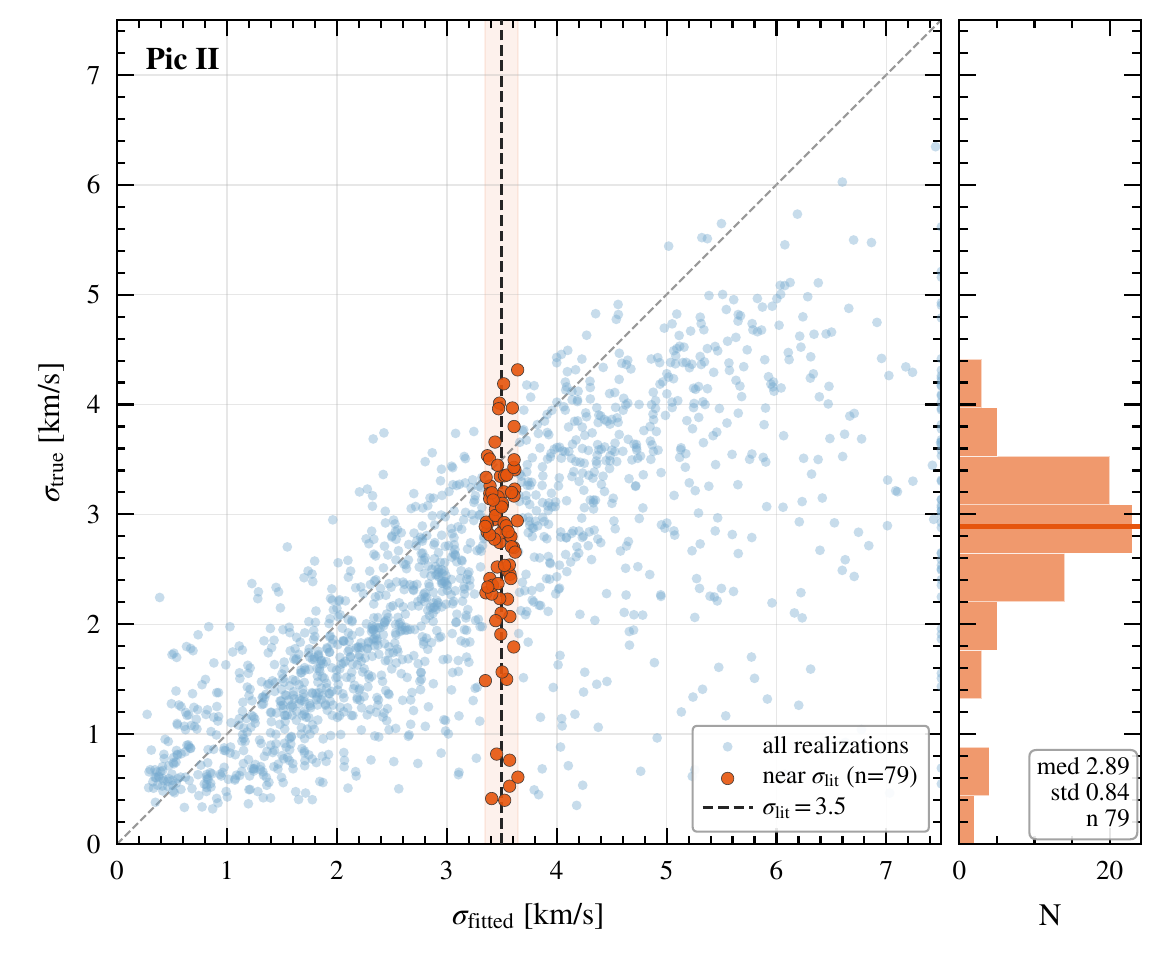}
    \includegraphics[width=0.48\linewidth]{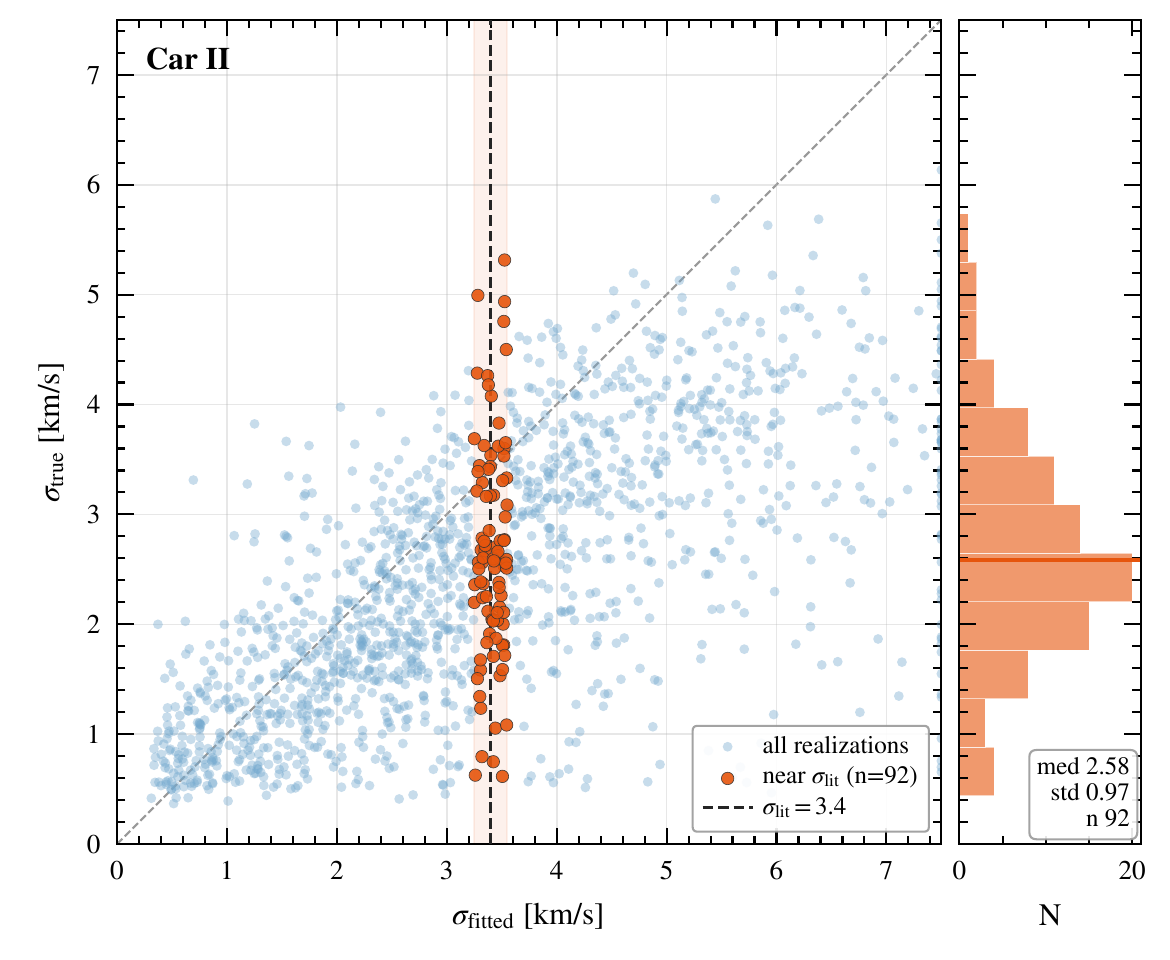}
    \caption{
    Empirical inversion of the published velocity dispersions for the four campaigns, at $f_{\rm b}=0.5$.
    Each blue point is one realization in which the intrinsic dispersion was drawn from a range and the campaign's real per-star schedule and cleaning were applied; the horizontal axis is the resulting cleaned dispersion and the vertical axis the intrinsic dispersion that produced it.
    The gray dashed line marks the unbiased case, and the systematic offset of the point cloud below it is the residual binary bias.
    Orange points are the realizations whose cleaned dispersion falls within $0.15$\,\kmsec\ of the published value (dashed vertical line); the side histogram shows the distribution of their intrinsic dispersions, whose median and scatter are the entries in Table~\ref{tab:ufds}.
    }
    \label{fig:ufd_inversion}
\end{figure*}

We perform similar experiments of empirical correction per system as shown in \bootes\,I. 
The intrinsic dispersion is drawn from a range, and we ask which true values produce a measured dispersion consistent with the published one under that campaign's schedule and cleaning. 
The inversion is shown in Figure~\ref{fig:ufd_inversion} and summarized in Table~\ref{tab:ufds}.

\begin{deluxetable*}{lcccccccc}
\tablecaption{Inversion of the published dispersions for four ultra-faint dwarf
campaigns, and the expected gain from one additional all-star follow-up epoch.
$\sigma_{\rm true}$ is the median true dispersion of the realizations whose cleaned
dispersion falls within $0.15$\,\kmsec{} of the published value, with $16$th--$84$th
percentiles, at $f_{\rm b}=0.5$. The final three columns describe the follow-up epoch:
$\Delta\sigma_{\rm proj}$ is the offset of the median recovered dispersion after the
follow-up from the published value, as annotated on Figure~\ref{fig:ufd_projection},
followed by the relative bias before and after the follow-up and the fraction of the
residual bias it removes.
\label{tab:ufds}}
\tablehead{
\colhead{System} & \colhead{$N_{\rm mem}$} & \colhead{Epochs} &
\colhead{Baseline} & \colhead{$\sigma_{\rm lit}$} & \colhead{$\sigma_{\rm true}$} &
\colhead{$\Delta\sigma_{\rm fitter}$} & \colhead{bias before / after} &
\colhead{bias removed} \\
\colhead{} & \colhead{} & \colhead{} & \colhead{(d)} &
\colhead{(\kmsec)} & \colhead{(\kmsec)} & \colhead{(\kmsec)} &
\colhead{} & \colhead{}
}
\startdata
Cen\,I   & 34 & 4 & 728 & $4.2$ & $3.79^{+0.36}_{-0.44}$ & $-0.08$ & $1.11 / 1.09$ & $18\%$ \\
Eri\,IV  & 28 & 4 & 113 & $6.1$ & $5.43^{+0.57}_{-0.55}$ & $-0.25$ & $1.13 / 1.08$ & $39\%$ \\
Pic\,II  & 13 & 5 & 747 & $3.5$ & $2.89^{+0.49}_{-0.80}$ & $-0.11$ & $1.21 / 1.17$ & $18\%$ \\
Car\,II  & 18 & 5 & 125 & $3.4$ & $2.58^{+0.99}_{-0.81}$ & $-0.21$ & $1.33 / 1.23$ & $28\%$ \\
\enddata
\end{deluxetable*}
\onecolumngrid

Table~\ref{tab:ufds} collects the inversion at $f_{\rm b}=0.5$. 
In every case the intrinsic dispersion implied by the mock lies below the published value: by $10\%$ for Cen\,I, $11\%$ for Eri\,IV, $17\%$ for Pic\,II and $24\%$ for Car\,II. 
The ordering tracks sample size rather than baseline: Pic\,II has the longest baseline of the four ($747$ days) but the smallest sample ($13$ members) and shows one of the largest offsets, while Eri\,IV recovers its published value comparatively well on a $113$-day baseline with $28$ members. 
This is the same behavior seen in the fiducial cases reinforces the conclusion of Section~\ref{sec:sparse_visit} that sample size and repeat coverage buy more than baseline alone.

The results are only weakly sensitive to the assumed binary fraction. 
Varying $f_{\rm b}$ from $0.3$ to $0.7$ shifts the inferred intrinsic dispersion by $0.15$\,\kmsec\ for Cen\,I, $0.22$ for Eri\,IV, $0.11$ for Pic\,II and $0.14$ for Car\,II. 
This insensitivity is the same one exploited in Section~\ref{sec:empirical_corr}, and it is what makes the inversion useful in practice: an observer need not know the binary fraction of their system to estimate the size of the correction.

\begin{figure*}
    \centering
    \includegraphics[width=0.48\linewidth]{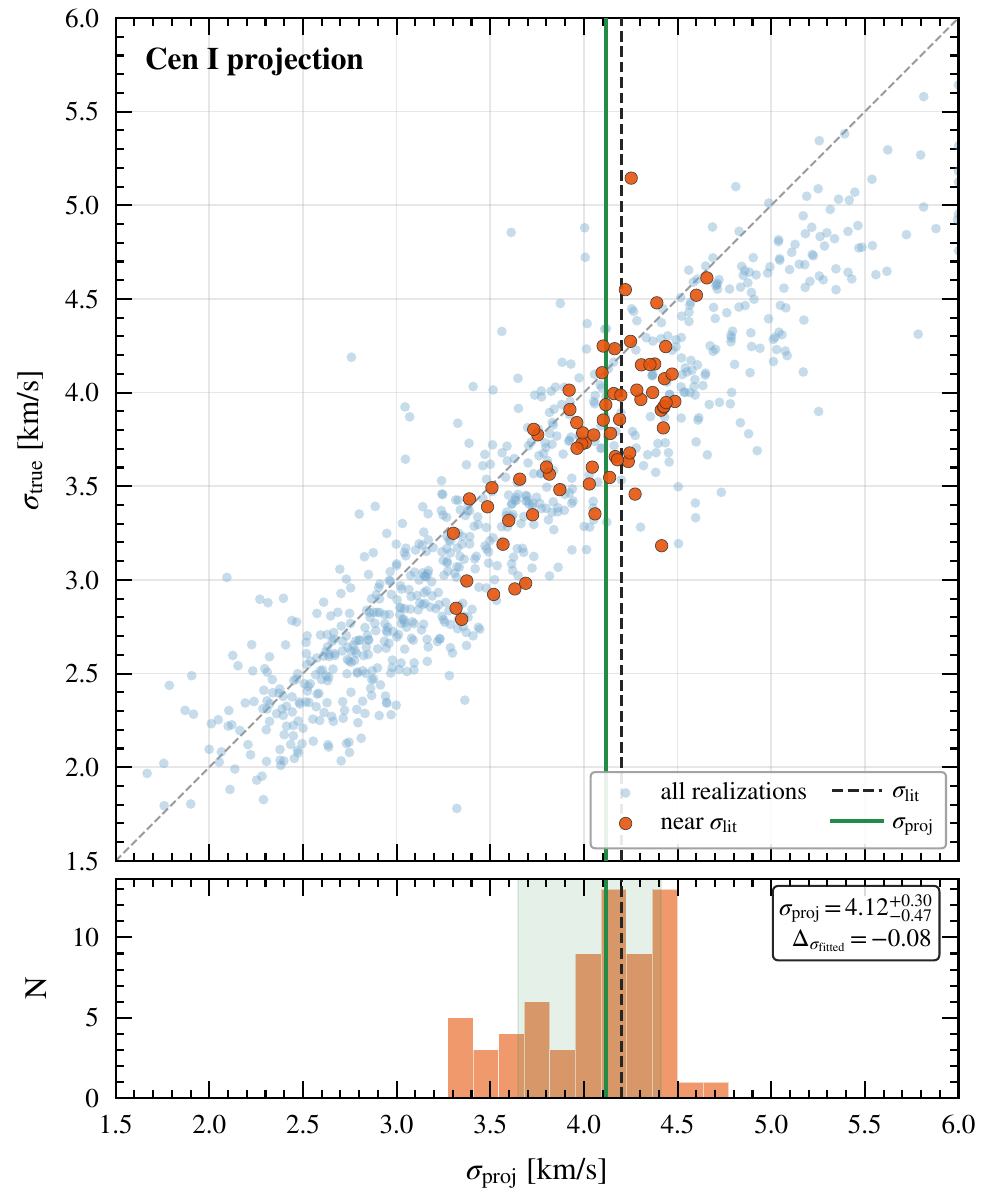}
    \includegraphics[width=0.48\linewidth]{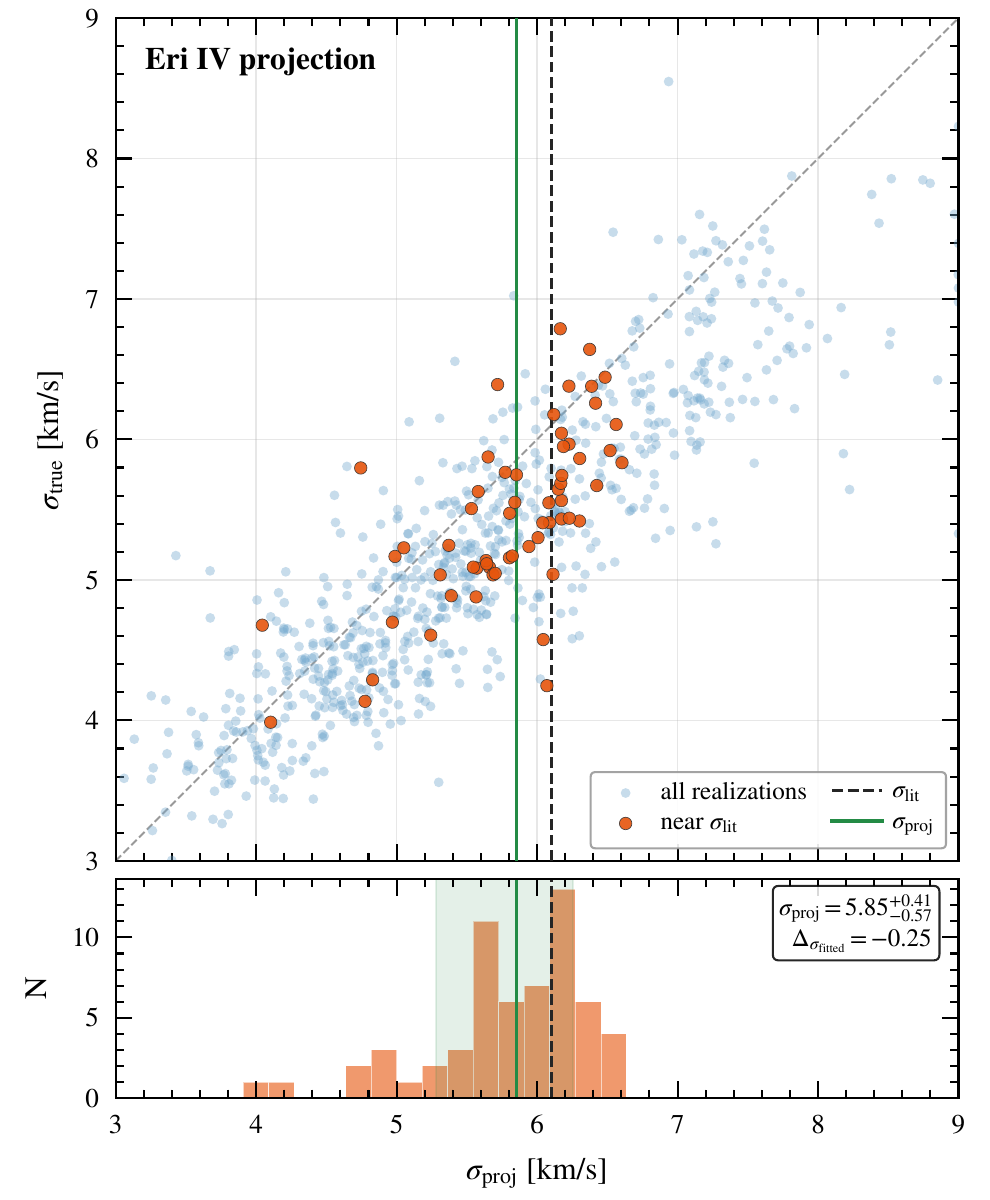} \\
    \includegraphics[width=0.48\linewidth]{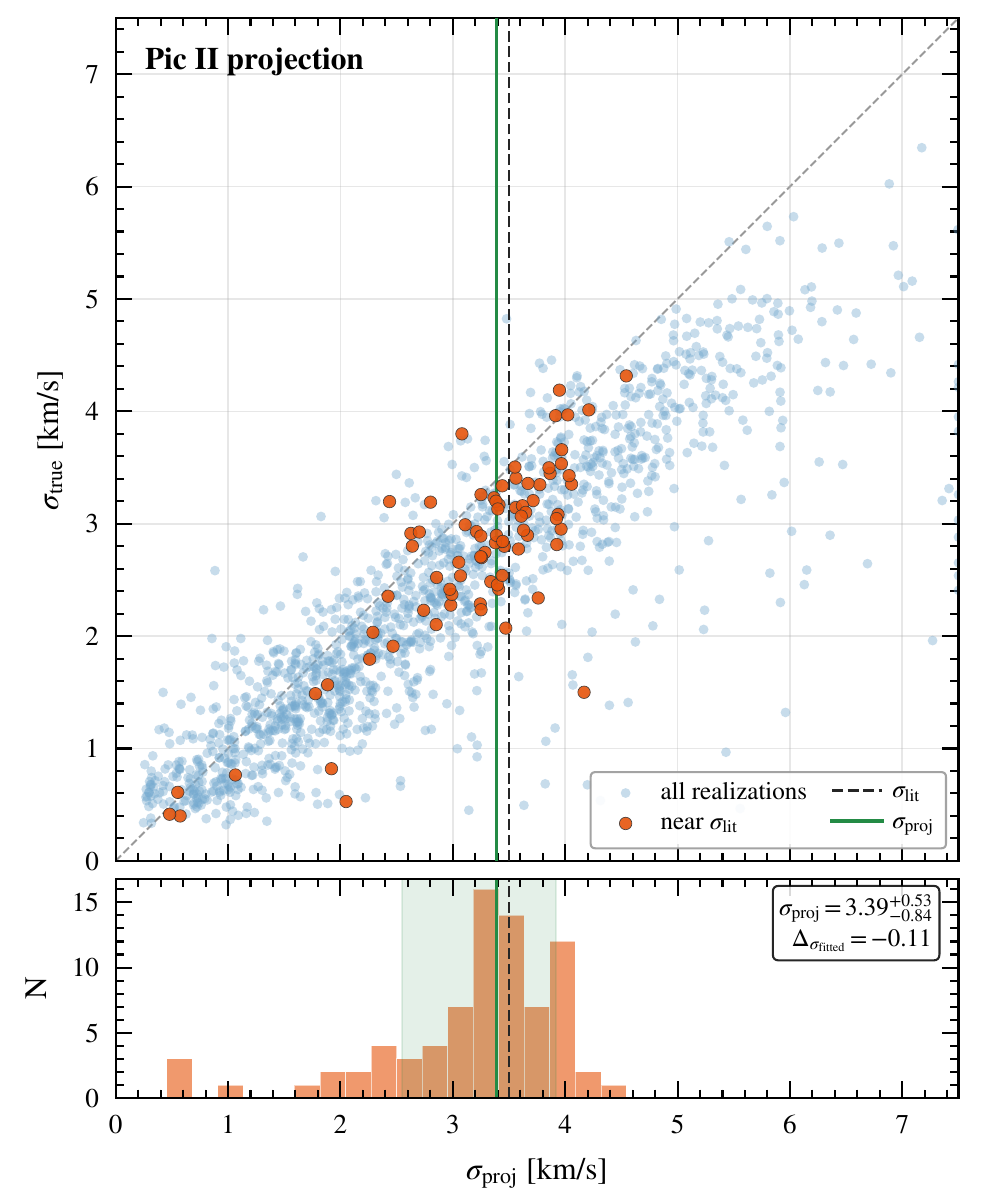}
    \includegraphics[width=0.48\linewidth]{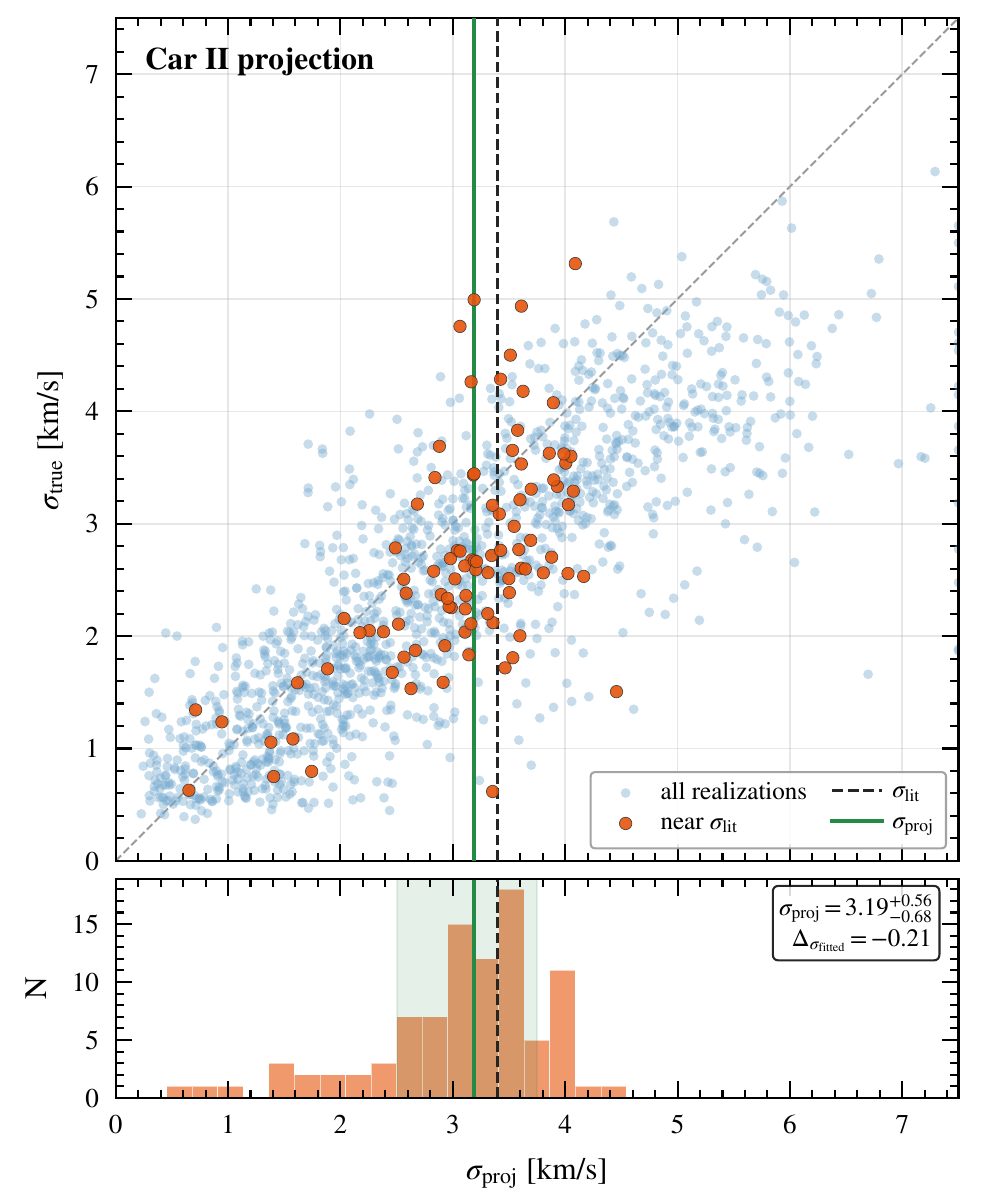}
    \caption{
    Effect of one additional all-star follow-up epoch on the four campaigns, at $f_{\rm b}=0.5$.
    Orange points are the same near-published realizations highlighted in Figure~\ref{fig:ufd_inversion}, now plotted against the cleaned dispersion recovered \emph{after} the follow-up epoch; blue points are all realizations.
    The black dashed line marks the published value and the green line the median of the near-published subset after the follow-up, so the separation between them is the expected downward shift.
    The lower panel of each pair shows the distribution of that shifted dispersion, with the median and its $16$th-$84$th percentile range annotated.
    }
    \label{fig:ufd_projection}
\end{figure*}

Each schedule also carries a hypothetical follow-up epoch, in which every member is re-observed at the last day of 2026 at its last recorded velocity uncertainty.
This lets us ask what a single additional night would gain.
Figure~\ref{fig:ufd_projection} shows the effect and the last three columns of Table~\ref{tab:ufds} quantify it.
The gain is real but partial: the follow-up removes between $18\%$ and $39\%$ of the residual bias, shifting the cleaned dispersion down by $0.08$-$0.25$\,\kmsec.
Car\,II and Eri\,IV gain most, and both are short-baseline campaigns ($125$ and $113$ days) in which a follow-up at the projected date more than doubles the total baseline; Cen\,I gains least, having already accumulated $728$ days.

The mechanism is worth separating, because it is not the one usually assumed.
For Pic\,II and Car\,II the follow-up flags no additional velocity variables at the median, yet still reduces the bias by $18\%$ and $28\%$ respectively.
The improvement there comes not from identifying new binaries but from averaging: a second measurement of a star whose first was taken at an unlucky orbital phase pulls its inverse-variance weighted velocity back toward the systemic value, shrinking the apparent dispersion even when the star is never recognized as a binary.
For the larger samples, Cen\,I and Eri\,IV, roughly one additional binary per realization is caught outright and both mechanisms contribute.
The practical implication is that a follow-up epoch is worth proposing even when the expected number of new binary detections is close to zero, and that its value is set by how much it extends the baseline relative to what already exists rather than by the number of stars it reclassifies.

We caution that these are forward-model inversions under an assumed binary population and a specific cleaning scheme, not re-reductions of the published data. 
The offsets in Table~\ref{tab:ufds} should be read as an estimate of the magnitude and sign of the residual binary bias in each campaign, not as revised measurements. 
The number of stars flagged as velocity variables in the mocks is small in every case ($0$-$4$ per realization at the final epoch), consistent with the small numbers of binaries identified in the original studies, and confirming that campaigns of this length and cadence identify only a modest fraction of the binaries present.

\end{CJK*}
\end{document}